%% file: bare_adv.tex
\documentclass[10pt,journal,compsoc]{IEEEtran}

\ifCLASSOPTIONcompsoc
  \usepackage[nocompress]{cite}
\else
  \usepackage{cite}
\fi
\ifCLASSINFOpdf
   \usepackage[pdftex]{graphicx}
   \graphicspath{{../pdf/}{../jpeg/}}
   \DeclareGraphicsExtensions{.pdf,.jpeg,.png}
\else
   \usepackage[dvips]{graphicx}
   \graphicspath{{../eps/}}
   \DeclareGraphicsExtensions{.eps}
\fi
\usepackage{url}
\usepackage{amsmath}
\ifCLASSINFOpdf
 \usepackage[pdftex]{thumbpdf}
\else
 \usepackage[dvips]{thumbpdf}
\fi
\usepackage{enumitem}
\usepackage{multirow}
\usepackage{amssymb}
\begin{document}
%
\title{HS-GCN: Hamming Spatial Graph Convolutional Networks for Recommendation}
%
%
%
%

\author{Han~Liu,~
        Yinwei~Wei,~\IEEEmembership{Member,~IEEE,}
        Jianhua~Yin,~\IEEEmembership{Member,~IEEE,}\\
        and~Liqiang~Nie,~\IEEEmembership{Senior~Member,~IEEE}
\IEEEcompsocitemizethanks{\IEEEcompsocthanksitem Han Liu, Jianhua Yin, and Liqiang Nie are with School of Computer Science and Technology, Shandong University, Qingdao 266200, China.\protect\\
E-mail: hanliu.sdu@gmail.com, jhyin@sdu.edu.cn, nieliqiang@gmail.com.
\IEEEcompsocthanksitem Yinwei Wei is with School of Computing, National University of Singapore, Singapore.\protect\\
E-mail: weiyinwei@hotmail.com.
\IEEEcompsocthanksitem Liqiang Nie is the corresponding author.}
}

%
%

\markboth{IEEE Transactions on Knowledge and Data Engineering}%
{Shell \MakeLowercase{\textit{et al.}}: Bare Advanced Demo of IEEEtran.cls for IEEE Computer Society Journals}
%



\IEEEtitleabstractindextext{%
\begin{abstract}
An efficient solution to the large-scale recommender system is to represent users and items as binary hash codes in the Hamming space. Towards this end, existing methods tend to code users by modeling their Hamming similarities with the items they historically interact with, which are termed as the first-order similarities in this work. Despite their efficiency, these methods suffer from the suboptimal representative capacity, since they forgo the correlation established by connecting multiple first-order similarities, i.e., the relation among the indirect instances, which could be defined as the high-order similarity. To tackle this drawback, we propose to model both the first- and the high-order similarities in the Hamming space through the user-item bipartite graph. Therefore, we develop a novel learning to hash framework, namely Hamming Spatial Graph Convolutional Networks (HS-GCN), which explicitly models the Hamming similarity and embeds it into the codes of users and items. Extensive experiments on three public benchmark datasets demonstrate that our proposed model significantly outperforms several state-of-the-art hashing models, and obtains performance comparable with the real-valued recommendation models.
\end{abstract}

\begin{IEEEkeywords}
Hashing, Efficient Recommendation, High-order Similarity, Graph Convolutional Network, Hamming Space.
\end{IEEEkeywords}}

\maketitle

\IEEEdisplaynontitleabstractindextext

%
\IEEEpeerreviewmaketitle

\ifCLASSOPTIONcompsoc
\IEEEraisesectionheading{\section{Introduction}\label{sec:introduction}}
\else
\section{Introduction}
\label{sec:introduction}
\fi

%
%
%
%
\IEEEPARstart{T}{he} recommender system is developed to locate the interested items from the overwhelming information according to users' preferences. Hence, how to measure the similarities between users and items is at the core of the personalized recommendation. Towards this end, existing studies~\cite{rendle2012bpr,wang2019neural,wei2019mmgcn} tend to follow a two-stage pipeline: representing the users and items with vectors~\cite{li2018deep}, and then predicting their interactions by measuring the similarities between vectors. Despite the remarkable performance, these methods still face an inevitable problem that the computation grows exponentially with increasing users and items~\cite{liu2018discrete}. Theoretically, for recommending top-$k$ preferred items for each user, the time complexity is $\mathcal{O}(NMK+NM\text{log}k)$ when there are $N$ users and $M$ items, represented by $K$-dimensional embeddings in the latent space.

Diving into these methods, we can easily find that the problem mainly comes from the user-item similarity computations~\cite{zhang2016discrete}. However, it is virtually impossible to design a new algorithm that not only can compute the similarity between two vectors but is more efficient than the inner-product. 
Therefore, some efforts have been dedicated to learning a new kind of representation---hash code---for the user and item, so as to alleviate the complexity~\cite{zhang2018discrete_KDD,wang2012semi}. In particular, benefiting from such a code consisting of $\pm 1$ bits, the measurement can be accelerated via XOR bit operation~\cite{weiss2009spectral}. Hence, the time complexity of recommendation is significantly reduced and even constant time search is made possible by exploiting lookup tables~\cite{zhang2016discrete}. Since the computation efficiency of the user-item similarity is supercharged, the large-scale recommendation could be conducted efficiently~\cite{kang2019candidate,lian2017discrete}, especially on mobile application where the computational resource is very limited~\cite{DBLP:conf/kdd/ChenYZHWW21,wang2020next}.
Nevertheless, since users and items are approximately represented as the binary vectors, the performance of hashing-based recommendation models tends to be suboptimal. To deal with this drawback, several methods are developed to enhance the representation ability of the hash codes~\cite{zhang2018discrete_WSDM,zhu2016deep,liu2016deep}. For instance, HashNet~\cite{cao2017hashnet} utilizes the deep neural networks to learn the user and item embeddings and then map them into the Hamming space to calculate the Hamming distance between two vectors. More recently, inspired by the success of graph convolutional networks in representation learning, HashGNN~\cite{tan2020learning} has been proposed to encode the local structural information into each node in the user-item interaction graph, and binarize their enhanced representations via a hash layer. 
\begin{figure}
    \centering
    \includegraphics[width=\linewidth]{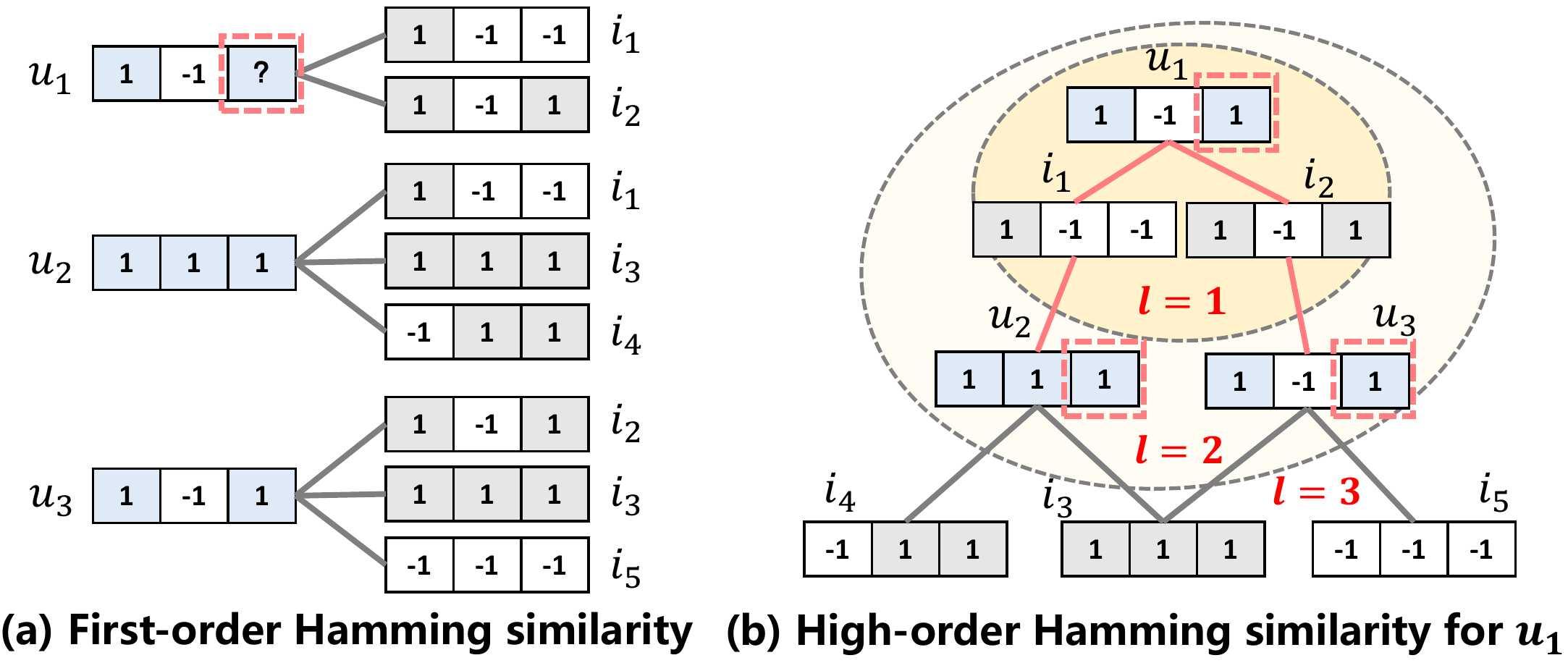}
    \vspace{-0.7cm}
    \caption{Illustration of the high-order and the first-order Hamming similarity in learning to hash.}
    \label{fig:intro}
    \vspace{-0.5cm}
\end{figure}

Although these methods enrich the user and item representations, we argue that they learn the hash codes merely with the similarities of interacted user-item pairs, but ignore the similarities hidden in the indirect interactions. To distinguish these similarities with the similarity computing in Euclidean space~\cite{wang2019neural}, we term them as the first-order and high-order Hamming similarities, respectively. To illustrate our argument, we connect the direct interactions to model the indirect interactions, which builds a user-item bipartite graph shown in Figure~$\ref{fig:intro}$(b), and compare it with the direct interactions shown in Figure~$\ref{fig:intro}$(a) on hash code learning. In particular, the users (\textit{i.e.}, $u_1$, $u_2$, and $u_3$) and items (\textit{i.e.}, $i_1$, $i_2$, and $i_3$) are represented by binary vectors, and the presence of an edge signifies that the user interacts with the item. To learn the hash codes for the users, both the deep learning based and GNN-based models code their interacted items and minimize their first-order Hamming similarities (\textit{i.e.}, $\langle u_1,i_1 \rangle$ and $\langle u_1,i_2 \rangle$), as shown in Figure~$\ref{fig:intro}$(a). However, for $u_1$, we argue that its code is hard to determine merely by the first-order Hamming similarity, because it lies right in the middle of two items. On the contrary, by incorporating the high-order Hamming similarity, as illustrated in Figure~$\ref{fig:intro}$(b), the hidden cues can be captured by measuring its high-order Hamming similarities, such as $\langle u_1,u_2 \rangle$ and $\langle u_1,u_3 \rangle$, largely facilitating the hash code learning.

Therefore, we propose to code the users and items by explicitly modeling the high-order Hamming similarity on the user-item bipartite graph. It is worth noticing that this is different with HashGNN that learns the nodes' representations in Euclidean space and then transforms them into hash codes. We argue that the supervision signal for the hash code optimization merely comes from the first-order interactions in the history, which may aggravate the information loss of high-order Hamming similarity during the transformation.

However, conducting the graph convolutional operations in the Hamming space to capture the Hamming similarity is non-trivial.  Following the terms in graph convolutional networks, we attribute the challenges into two aspects:
\begin{itemize}[leftmargin=*]
    \item Different from the operations on the continuous vectors, each element in hash codes is restricted to the binary values. Therefore, for each node, \textit{how to aggregate the information from its neighbors in the Hamming space is the first challenge we face}.
    \item In order to represent the nodes in Hamming space, it is essential to preserve their Hamming similarities with the codes of their neighbors. Hence, \textit{how to explicitly encode the Hamming similarities into each node is another technical challenge in this work}.
\end{itemize}

In order to address the outlined challenges, we develop a novel hashing-based recommendation model, named Hamming Spatial Graph Convolutional Networks (\textbf{HS-GCN}), consisting of the initial, propagation, and prediction layers.  
Specifically, in the initial layer, we recognize the user-item interactions as a bipartite graph and initialize the nodes with a hash code generation method proposed in~\cite{cao2017hashnet}, which guarantees the feasibility of end-to-end optimization. Upon the constructed graph, we devise a code propagation layer to implement the graph convolutional operations in Hamming space. 
More specifically, the layer could be divided into two components --- hash code aggregation and hash code encoding. The former aggregates information from first-order similar neighbors by counting the bit-wise signs (\textit{i.e.}, finding out the dominated bit value in each dimension) of their corresponding hash codes. And the latter injects the Hamming similarities into the node hash codes by refining their bits consistent with the bit-wise signs of their neighbors. 
With the stacked propagation layers, we iteratively embed the learned hash codes into the nodes. And then, the interactions of user-item pairs could be predicted by scoring their affinities at the prediction layer. To evaluate our proposed model, we conduct extensive experiments on three publicly accessible datasets and compare the performance with several state-of-the-art baselines. In addition to outperforming the hash-based models (e.g., HashNet and HashGNN), our proposed method achieves the results comparable with the real-valued based models, such as PinSage and GraphSAGE.


Overall, the main contributions of our work are summarized in three-folds:
\begin{itemize}[leftmargin=*]
\item To the best of our knowledge, this is the first attempt to explicitly model the high-order Hamming similarity between users and items in the recommender system, which enhances the representation ability of hash codes and accordingly optimizes the user-item interaction prediction. 
\item We present a GCN-based hashing recommendation model, named Hamming Spatial Graph Convolutional Network (HS-GCN), which explicitly captures the first- and high-order Hamming similarities. In particular, we develop the novel graph convolutions operation---hash code aggregation and hash code encoding---in the Hamming space.
\item Extensive experimental results on three real-world datasets have demonstrated that the proposed model yields a substantial performance improvement compared with several state-of-the-art baselines. As a side contribution, we have released the codes to facilitate other researchers: \url{https://github.com/hanliu95/HS-GCN}.
\end{itemize}


 



\input{5_related-work}
\input{2_model}

\input{3_experiment}

\input{4_results}

\input{6_conclusion}

\ifCLASSOPTIONcompsoc
  \section*{Acknowledgments}
  This work is supported by the National Natural Science Foundation of China, No.:U1936203; the Key R\&D Program of Shandong (Major scientific and technological innovation projects), No.:2020CXGC010111; and Beijing Academy of Artificial Intelligence(BAAI).
\else
  \section*{Acknowledgment}
\fi


\ifCLASSOPTIONcaptionsoff
  \newpage
\fi



\bibliographystyle{IEEEtran}
\bibliography{reference}
\end{document}

%% file: 5_related-work.tex
\section{Related Work}
In this section, we briefly review graph-based methods for recommendation and learning to hash for recommendation.

\subsection{Graph-based Methods for Recommendation}
Machine learning on graphs is an important task with the advantage of incorporating structural information. As one of the primary applications, representation learning on the user-item graph structure has been widely-studied in recommendation scenarios by utilizing information propagation. In this line of research, ItemRank~\cite{gori2007itemrank} and BiRank~\cite{he2016birank} make early efforts for label propagation. In particular, these methods directly propagate the user preference scores (\emph{i.e.}, labels) on the graph, scoring connected items with similar labels for a user. 
However, these methods are essentially neighbor-based and insufficient to encode structural information of a graph. 

Recently, GNNs have received focused attention~\cite{HUIGN,GRCN}, since GNNs have special advantage on modeling the graph structure, especially information propagation, to guide the representation learning.
However, early GNN-based methods suffer from expensive computation costs as the graph convolution on the spectral domain. Subsequently, 
GCNs exploit a graph convolution operation on the spatial domain, which aggregates the embeddings of neighbors to refine the embedding of the target node~\cite{wei2019mmgcn,liu2021interest}. Motivated by the efficiency of graph convolution, GC-MC~\cite{berg2017graph}, PinSage~\cite{ying2018graph}, and NGCF~\cite{wang2019neural} employ GCNs to capture the propagation of latent Collaborative Filtering (CF) signals in the user-item interaction graph for recommendation. However, later studies argue the excessive complexity of GCNs. For example, LightGCN~\cite{he2020lightgcn} develops a linear model by removing all redundant parameters. Experimental results demonstrate that the simplified design outperforms original GCNs by a large margin. This shows that the nonlinearity of GCNs is unnecessary for CF recommendation. 

As aforementioned, we summarized two key points: 1) graph-based methods shed light on modeling the relational information propagation like the high-order similarity in the Hamming space, based on a user-item graph; 2) graph convolution has better performance after being simplified, verifying that the nonlinearity is not necessary in graph-based recommendation. Our work is highly inspired by these analyses, since the high-order Hamming similarity can be effectively modeled by the graph-based mechanism, and all the operations are simply linear in the Hamming space. Moreover, Heterogeneous GNN is proposed to be adapted upon structurally complex graphs to utilize richer information within them. Based on this technology, some methods effectively learn representations by aggregating different types of neighboring information in heterogeneous information network, such as HERec~\cite{shi2018heterogeneous}, MCRec~\cite{hu2018leveraging}, MEIRec~\cite{fan2019metapath}, and ie-HGCN~\cite{yang2021interpretable}. On contrary, our method works with simple interaction information which can be easily collected.

\subsection{Learning to Hash for Recommendation}
Another relevant research line is learning to hash for recommendation, which proceeds along two directions: unsupervised hashing and supervised hashing~\cite{li2020weakly}. The former~\cite{jegou2010product,liu2011hashing} learns hash functions that encode objects to binary codes by training from unlabeled data, while the latter~\cite{kulis2009learning,liu2012supervised,norouzi2011minimal,shen2015supervised} aims to learn more discriminative hash codes by exploring labeled signals, such as the feedback between users and items in recommendation scenarios. In general, early learning to hash for recommendation methods are essentially two-stage approaches~\cite{gong2012iterative}.
However, the Hamming similarity between learned hash codes might not correspond to the original relevance between a user and an item, since there are quantization deviations when thresholding real values to binary bits during the binarization step. 
DCF~\cite{zhang2016discrete} tackles the challenging discrete optimization problem and learns user and item hash codes directly. Therefore, the learned hash codes are able to model the intrinsic user-item relevance.

A prevailing trend is to leverage deep learning for recommendation. For instance, NFM~\cite{he2017nfm} successfully introduces deep learning to enhance representation learning and matching function modeling in recommendation. In the light of this, deep learning to hash is subsequently developed and yields promising recommendation performance. Early efforts of deep learning to hash also adopt a two-stage strategy: the first stage employs deep networks to learn continuous representations and the second one uses the $\text{sign}(\mathbf{x})$ function to binarize the learned representations into binary hash codes. This category of methods, such as CNNH~\cite{xia2014supervised}, DNNH~\cite{lai2015simultaneous}, and DHN~\cite{zhu2016deep}, also suffer from the quantization deviation. To alleviate it, HashNet~\cite{cao2017hashnet} is proposed to devise a one-stage learning to hash technique to decrease the quantization deviation of binarization. It approximates the $\text{sign}(\mathbf{x})$ function with function $\text{tanh}(\beta\mathbf{x})$, where $\beta$ is a scaled parameter that increases during training. The infinite approximation makes the deviation negligible, and thus contributes to better recommendation performance.

In a sense, the aforementioned hashing techniques only consider the first-order similarity between hash codes. Therefore, we introduced graph-based techniques in hash learning to capture the high-order Hamming similarity. It is worth mentioning that several recent efforts have incorporated GNN insights into hashing, such as DGCN-BinCF~\cite{wang2019binarized}, GCNH~\cite{zhou2018graph}, and HashGNN~\cite{tan2020learning}. Particularly, HashGNN~\cite{tan2020learning} sets up a GNN in the continuous space, followed by a straight through estimator~\cite{bengio2013estimating} for generating hash codes in the Hamming space. However, the hashing step leads to information loss, which impedes the capturing of intrinsic high-order Hamming similarity. To bridge this gap, we directly constructed a GCN in the Hamming space to model the high-order similarity.

Different from the diffusion models~\cite{zhou2012fusion,yang2019efficient,wang2014similarity} which combine different similarity measures by multiple similarity graphs~\cite{liu2017cross,bai2017ensemble,wang2021dissecting,jiang2019data}, our method extends the Hamming similarity from first-order to high-order via a single graph, and the high-order similarity is exploited to learn hash codes with better representative capacity. Also different from hypergraph convolution networks that exploit multi-modal data in hypergraph for representation learning~\cite{feng2019hypergraph}, our method is simply based on the single-modal interaction information in user-item graph, which can be obtained more easily.

%% file: 2_model.tex
\section{Preliminaries}
We first give the definition of Hamming similarity, a similarity measurement for the hash codes in the Hamming space, highlighting that it is related to the number of the same bits. We then formulate the problem to be solved in our work.
\subsection{Hamming Similarity}
In the Hamming space, the user-item similarity is equivalent to the similarity between their corresponding hash codes, called Hamming similarity. 
Given user $u$ and item $i$, we denote their hash codes as  $\mathbf{h}_u\in\{\pm 1\}^K$ and $\mathbf{h}_i\in\{\pm 1\}^K$ respectively, where $K$ represents the length of the codes. Accordingly, the Hamming similarity between them is defined as:
\begin{equation}
    \text{sim}(u,i)=\frac{1}{K}\sum_{k=1}^K\mathbb{I}(h_{uk}=h_{ik}),
\end{equation}
where $h_{uk}$ and $h_{ik}$ are the $k$-th bits of $\mathbf{h}_u$ and $\mathbf{h}_i$, respectively, and $\mathbb{I}(\cdot)$ denotes the indicator function that returns $1$ if the statement is true and $0$ otherwise. Furthermore, it could be proven that $\text{sim}(u,i)$ is proportional to the number of same bits in $\mathbf{h}_u$ and $\mathbf{h}_i$. As such, we rewrite the formulation~\cite{zhang2016discrete} as:
\begin{equation}
    \text{sim}(u,i)=\frac{1}{2} + \frac{1}{2K}{\mathbf{h}_u}^{\top}\mathbf{h}_i\propto {\mathbf{h}_u}^{\top}\mathbf{h}_i,
\end{equation}
where the derivation process is omitted. Based on this formulation, we directly employ the inner product of  ${\mathbf{h}_u}$ and $\mathbf{h}_i$ to measure the similarity between $u$ and $i$.

\subsection{Problem Formulation}
In this work, we aim to tackle the problem of mapping nodes in a user-item graph to hash codes for recommendation. We formulate the problem to be addressed in this paper as the following:
\begin{itemize}[leftmargin=*]
\item\textbf{Input:} the user-item bipartite graph  $\mathcal{G}=(\mathcal{V},\mathcal{E})$ that records historical user-item interactions in a graph structure. Whereinto, the node set consists of the user set $\mathcal{U}$ of $N$ users and the item set $\mathcal{I}$ of $M$ items, formally $\mathcal{V}=\mathcal{U}\cup\mathcal{I}$. And, the edge set $\mathcal{E}$ contains the observed user-item interactions in the training phase.
\item\textbf{Output:} the hash codes of all users and items, defined as $\{\mathbf{h}_u^{*}|u\in\mathcal{U}\}\cup\{\mathbf{h}_i^{*}|i\in\mathcal{I}\}$. Since ${\mathbf{h}_u^*}^{\top}\mathbf{h}_i^*$ denotes the similarity score between user $u$ and item $i$, and on top of that we can generate a ranking list of items for a given user and hence solve the practical problem of recommendation. To accurately measure the similarity score, we focus on devising a GCN for hash code propagation, which is able to learn node hash codes with high-order Hamming similarities latent in the user-item bipartite graph.
\end{itemize}
\begin{figure}[t]
\centering
\includegraphics[width=\linewidth]{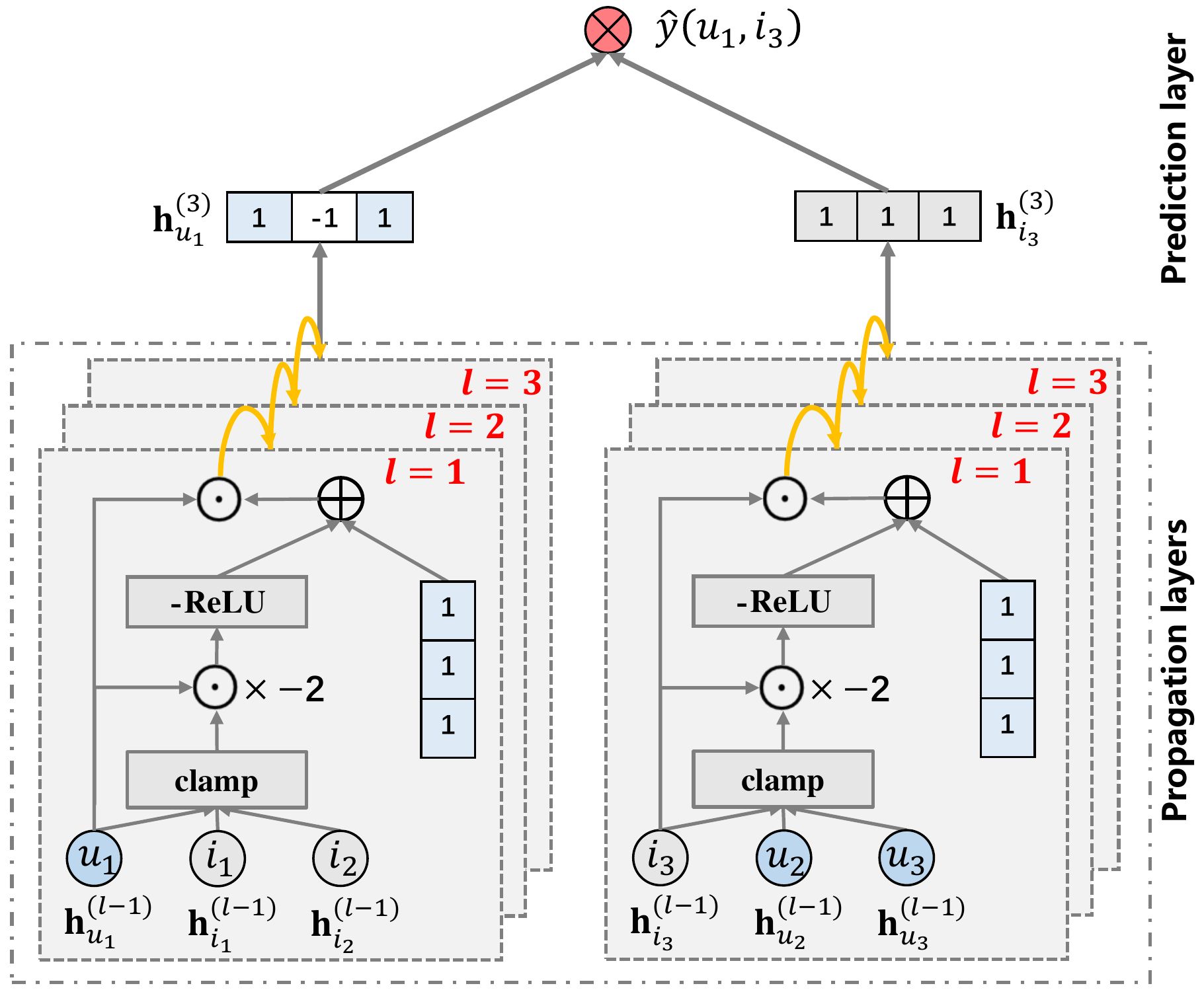}
\vspace{-0.75cm}
\caption{Illustration of our proposed HS-GCN model (the arrow lines present the flow of information). The hash codes of user $u_1$ (left) and item $i_3$ (right) are refined with multiple hash code propagation layers.}
\label{fig:architecture}
\end{figure}

\section{Methodology}
In this section, we detail our proposed HS-GCN model, as illustrated in Figure~$\ref{fig:architecture}$. Specifically, the model consists of three components: 1) the initial layer that yields the trainable hash codes for the nodes in the user-item bipartite graph; 2) the propagation layer that explicitly model the Hamming similarities of node pairs and inject them into their codes for enhancement of the representation ability; and 3) the prediction layer that scores the similarity between the user and item by conducting the inner product of their hash codes. Moreover, we provide the optimization of HS-GCN in the form of matrix and compare HS-GCN with a graph-based learning to hash method HashGNN~\cite{tan2020learning}. 


\subsection{Initial Layer}
The initial layer serves to generate the initial hash codes $\mathbf{h}_u \in \{\pm 1\}^K$ ($\mathbf{h}_i \in \{\pm 1\}^K$) for user $u$ (item $i$), where $K$ is the size of hash codes. Obviously,
 due to its discrete value, the hash code cannot be trained by the gradient descent. Hence, we alternatively employ a proxy to represent the users and items during the training phase. 
 To achieve this goal, we follow the existing work~\cite{cao2017hashnet} to reformulate the hash codes as, 
 \begin{equation}
     \mathbf{h}_u=\lim_{\beta\rightarrow\infty} \tanh(\beta\mathbf{e}_u),
     \mathbf{h}_i=\lim_{\beta\rightarrow\infty} \tanh(\beta\mathbf{e}_i),
\end{equation}
where $\mathbf{e}_u\in\mathbb{R}^K$ and $\mathbf{e}_i\in\mathbb{R}^K$ denote the parameter vector of user $u$ and item $i$, respectively. By scaling the value of $\beta$, the outputs can approximate the desired codes. With the help of such a proxy, it is capable of addressing the ill-posed gradient issue and optimizing the representations during the training process. 
When the model reaches the convergence, we could infer the interactions between users and items via the learned hash codes, instead of their approximations. 

In what follows, we take the approximate codes of users and items as the initial representations of the nodes in the user-item graph and detail our devised graph convolutional operations in Hamming space. For notational convenience, we denote them as $\mathbf{h}^{(0)}_u$ and $\mathbf{h}^{(0)}_i$, identifying the nodes' codes at the $0$-th layer.

\subsection{Propagation Layer}
To perform the graph convolutional operations in Hamming space, we devise a novel propagation layer, which disentangles the operations into the hash code aggregation and hash code encoding. With the stacked propagation layers, we could explicitly model the first- and high-order Hamming similarities hidden in the user-item graph and optimize the nodes' representations by injecting the learned Hamming similarities into the codes.

In this section, without particular clarification, we elaborate our proposed model on the user nodes and conduct the same operations on item nodes.

\subsubsection{Hash Code Aggregation}
Regarding the graph convolutional operations, it is essential to capture the local structure information for each centric-node. 
To the end, we are inspired by the prior studies~\cite{wang2019neural,wei2019mmgcn} and develop a new aggregation to explicitly model the local structure information in Hamming space, which reveals the first-order Hamming similarity. Formally, the aggregation at $(l+1)$-th propagation layer can be defined as, 
\begin{equation}
    \mathbf{m}_{u}^{(l)}=f\big(\mathbf{h}_u^{(l)}, \{\mathbf{h}_i^{(l)}|i\in \mathcal{N}_u\}\big),
\end{equation}
where $\mathbf{h}_u^{(l)}\in \{\pm 1\}^K$ and $\mathbf{h}_i^{(l)}\in \{\pm 1\}^K$ recursively denote the hash codes of user $u$ and item $i$ at the $l$-th layer propagation, respectively. And, $\mathcal{N}_u$ denotes the set of item neighbors of user $u$, which $u$ directly interacted with. Using the function $f(\cdot)$, we could capture the informative signal $ \mathbf{m}_{u}^{(l)}$ distilled by incorporating the node with its neighbors. 


However, the implementation of this function is not straightforward in Hamming space. In this work, we implement it in two steps. In particular, considering the fact that the transformation of the weighted matrix is inapplicable in Hamming space, we first adopt the non-weighted sum to integrate the bit-wise signals of the hash codes. Accordingly, we rewrite the aggregation in Hamming space:
\begin{equation}\label{eq:agg}
    \mathbf{m}_{u}^{(l)}=f(\mathbf{h}_u^{(l)}+\sum_{i\in \mathcal{N}_u}\mathbf{h}_i^{(l)}).
\end{equation}
Then, since the hash codes constrain binary values~(\textit{i.e.,} +1 and -1), we capture the bit-wise dominate value in the hash codes of centric node and its neighbors, which is different with the graph operations in Euclidean space, like mean- and max-pooling. We attribute it to the binomial distribution of the hash code. 
As such, we aim at designing a bit-wise function $f(\cdot)$ for aggregated codes and use each bit in $\mathbf{m}_u^{(l)}$ as the sign to reflect the dominate value at the corresponding bit of codes of $u$ associated with its neighbors. 
Thereby, jointly considering the continuity of function, we implement $f(\cdot)$ with the interval limited clamp function as follows:
\begin{equation}
    \mathbf{m}_{u}^{(l)}=\text{clamp}(\mathbf{x}) =\left\{
\begin{aligned}
1 &,~\text{if}~\mathbf{x} > 1 \\\
\mathbf{x} &,~\text{if}~-1\le \mathbf{x} \le 1\\\
-1 &,~\text{if}~\mathbf{x} < -1
\end{aligned}
\right.
,
\end{equation}
where $\mathbf{x}=(\mathbf{h}_u^{(l)}+\sum_{i\in \mathcal{N}_u}\mathbf{h}_i^{(l)})$. 
Owing to the characteristic of the clamp function, we can guarantee 
$\mathbf{m}_{u}^{(l)} \in \{-1, 0, 1\}^K$.
For each node, the function is expected to reflect the distribution of its neighbors' hash codes. It is able to determine the value of centric node, making it similar with its neighbors. Taking one bit in hash code as an example, the function will output $-1$ on this bit  when the majority of corresponding bits in neighbors' hash codes are $-1$, and vice versa. In a nutshell, we propose an aggregation operation in Hamming space, which integrates the codes from the centric node associated with its neighbors and distills the signal to reflect the bit-wise value distributions of hash codes.

\subsubsection{Hash Code Encoding}
Obtaining the aggregated codes from the local structure, we then encode the information into the nodes to preserve their first-order Hamming similarities towards their connected nodes. 
As for each node, it is of importance to minimize the total of bits which are different from the corresponding bits of its neighbors' codes. 
Therefore, we first compare the code of user $u$~(\textit{i.e.}, $\mathbf{h}_u^{(l)}$) with the obtained signs~(\textit{i.e.}, $ \mathbf{m}_{u}^{(l)}$) to capture the number of different bits, as, 
\begin{equation}
    \mathbf{d}_u^{(l)}=\mathbf{h}_u^{(l)}\odot \mathbf{m}_{u}^{(l)},
\end{equation}
where $\odot$ denotes the element-wise product, and $\mathbf{d}_u^{(l)}\in \{-1, 0, 1\}^K$ is a vector consisting of the bit-wise comparison result. More specifically, the value $-1$ in the output indicates the corresponding bit differs from most of neighbors'; whereas, the value $1$ means the node has the same bit value with the majority of nodes around it. 
Besides, when the value equals to $0$, it implies that the corresponding bit is caught in the middle.

Then, we could refine the codes of user $u$ according to each value in the obtained vector. 
Considering the element $0$ existing in the vector, this operation is hardly implemented by conducting the element-wise product of two vectors, which probably outputs zero in addition to the binary value~(\textit{i.e.,} $+1$ and $-1$). 
Therefore, we design a new function to avoid the negative influence of the value $0$. It is formally defined as,  
\begin{equation}
    \mathbf{c}_u^{(l)}=-\text{ReLU}(-2 \times \mathbf{d}_u^{(l)})+ones,
\end{equation}
where $ones=\{1\}^K$ is the vector that adjusts the outputs following the binary constraint. Based on this transform function, the values are tuned within $\{-1, 1\}$. Note that we convert the values $0$ to $1$ rather than $-1$, making the $-1$ values in $ \mathbf{c}_u^{(l)}$ exactly indicate the unquestioned bit differences between the node and its neighbors. This will improve the reliability of the following bit-wise refinement.

With the tuned vector, we could
refine the hash code $\mathbf{h}_u^{(l)}$ by maximizing the similarity with its neighbors. 
Since $\mathbf{c}_u^{(l)}$ indicates the bits which are distinct with the majority of its neighbors, we utilize $\mathbf{c}_u^{(l)}$ to refine the bits of $\mathbf{h}_u^{(l)}$ with the element-wise product, formally, 
\begin{equation}\label{eq:ref}
    \mathbf{h}_u^{(l+1)}=\mathbf{c}_u^{(l)}\odot \mathbf{h}_u^{(l)},
\end{equation}
where $\mathbf{h}_u^{(l+1)}$ denotes the hash code of user $u$ learned at the ($l$+1)-th propagation layer. 
Analogously, we can obtain the hash code $\mathbf{h}_i^{(l+1)}$ of item $i$ by propagating hash codes from itself and its connected users. In summary, the designed hash code propagation layer contributes to exploit the first-order similarity information to relate user and item hash codes in the Hamming space.

\subsubsection{Hash Code Propagation Rule in the Matrix Form} To offer a holistic view of the hash code propagation architecture and facilitate the batch implementation, we provide the matrix form of the layer-wise propagation rule as:
\begin{equation}
\begin{split}
    \mathbf{H}^{(l+1)}=
    \Big(&-\text{ReLU}\big(-2\times\text{clamp}\big((\mathbf{A}+\mathbf{I})\mathbf{H}^{(l)}\big)\odot\mathbf{H}^{(l)}\big)\\&+Ones\Big)\odot\mathbf{H}^{(l)},
\end{split}
\end{equation}
where $\mathbf{H}^{(l)}\in\{\pm 1\}^{(N+M)\times K}$ are the hash codes for users and items obtained after $l$ steps of propagation. $\mathbf{H}^{(0)}$ is the matrix of the initial hash codes after the initial layer, where $\mathbf{h}_u^{(0)}=\mathbf{h}_u$ and $\mathbf{h}_i^{(0)}=\mathbf{h}_i$. $\mathbf{I}$ denotes the identity matrix, and $Ones\in \{1\}^{(N+M)\times K}$ is a matrix of all ones. $\mathbf{A}$ denotes the adjacency matrix for the user-item graph, which is formulated as:
\begin{equation}
    \mathbf{A}=
    \left[
\begin{array}{cc}
\mathbf{0} & \mathbf{R} \\
\mathbf{R}^{\top} & \mathbf{0} 
\end{array}
\right],
\end{equation}
where $\mathbf{R}\in\{0,1\}^{N\times M}$ is the user-item interaction matrix, and $\mathbf{0}$ is a all zero matrix. By implementing the matrix form propagation rule, we can simultaneously update the hash codes for all users and items in a rather efficient way. Moreover, it assists us to discard the node sampling procedure, which is commonly used to make graph convolution networks applicable on large-scale graph.

\subsection{Prediction Layer}
We stack multiple propagation layers to explore the high-order similarities between hash codes, which are vital for upgrading the hash codes to finally estimate the relevance between the user-item pair. In particular, we define $L$ as the depth of the final propagation layer, and the hash code of user $u$ is recursively defined as:
\begin{equation}\label{eq:m_l}
\left\{
\begin{aligned}
    &\mathbf{h}_u^{(1)}=\mathbf{c}_u^{(0)}\odot\mathbf{h}_u,\\\
    &\mathbf{h}_u^{(2)}=\mathbf{c}_u^{(1)}\odot\mathbf{h}_u^{(1)}=\mathbf{c}_u^{(1)}\odot\mathbf{c}_u^{(0)}\odot\mathbf{h}_u,\\\
    &\cdots\cdots\\\
    &\mathbf{h}_u^{(L)}=\mathbf{c}_u^{(L-1)}\odot\mathbf{h}_u^{(L-1)}=\mathbf{c}_u^{(L-1)}\odot\cdots\odot \mathbf{c}_u^{(0)}\odot \mathbf{h}_u.
\end{aligned}
\right.
\end{equation}
It shows that our model obtains the final hash codes by element-wise multiplying the initial ones with a series of refinement vectors. These refinement vectors then modulate the bit signals of user $u$'s hash code layer by layer. As thus, the high-order Hamming similarity is injected into the hash code learning process. Note that the length of hash code is identically $K$ in each layer 
according to Eqn.(\ref{eq:m_l}). Analogously, we can obtain the hash code for item $i$ at the $L$-th layer.


After performing $L$ propagation layers, we obtain multiple hash codes for user $u$, termed $\{\mathbf{h}_u^{(1)},\cdots,\mathbf{h}_u^{(L)}\}$. It is worth noting that we choose not to concatenate them as the final representation for a user, which is widely used in the graph-based recommendation models~\cite{wang2019neural,he2020lightgcn}. The reason is that the length of concatenation codes probably harms the computational efficiency and memory space. Alternately, we take the hash codes at the last layer~(\textit{i.e.,} $\mathbf{h}_u^{(L)}$ and $\mathbf{h}_i^{(L)}$) as the representations of user $u$ and item $i$ to predict their interaction, formally, 
\begin{equation}
    \mathbf{h}_u^*=\mathbf{h}_u^{(L)},
    \quad
    \mathbf{h}_i^*=\mathbf{h}_i^{(L)}.
\end{equation}

Finally, we apply the inner product to estimate the similarity between hash codes of the target user and item, which can be equivalently treated as the user's preference towards the item:
\begin{equation}
    \hat{y}_{ui}={\mathbf{h}_u^*}^{\top}\mathbf{h}_i^*.
\end{equation}

\subsection{Optimization}
In this work, we optimize the proposed model based on the users' implicit feedback, such as observations and purchases~\cite{CLCrec,Yu}.  
Compared to explicit ratings, the implicit feedback is easier to collect but more challenging for exploring the user preference, due to its scarcity of negative feedback. 
To optimize the model by the maximization likelihood estimation~(MLE), we formulate the likelihood function $P(r_{ui}|\mathbf{h}_u^*,\mathbf{h}_i^*)$ as the probability of the interaction between user $u$ and item $i$ equals $r_{ui}$ given their trainable codes $\mathbf{h}_u^*$ and $\mathbf{h}_i^*$. As such, its optimal objective can be instantiated as the pairwise logistic function as follows:
\begin{equation}\label{eq:P}
    P(r_{ui}|\mathbf{h}_u^*,\mathbf{h}_i^*)=\left\{
    \begin{aligned}
    &\sigma(\hat{y}_{ui}),~r_{ui}=1,\\
    &1-\sigma(\hat{y}_{ui}),~r_{ui}=0,
    \end{aligned}
    \right.
\end{equation}
where $\sigma(\cdot)$ is the standard sigmoid function. And, $\hat{y}_{ui}={\mathbf{h}_u^*}^{\top}\mathbf{h}_i^*$ is the estimated Hamming similarity between $\mathbf{h}_u^*$ and $\mathbf{h}_i^*$. In this case, the more similar the hash codes are, the larger the conditional probability $P(1|\mathbf{h}_u^*,\mathbf{h}_i^*)$ will be, and vice versa. Therefore, Eqn.($\ref{eq:P}$) is a reasonable extension of the logistic regression classifier for the pairwise classification problem based on hash codes. Further, we can optimize the parameter by minimizing the following loss function:
\begin{equation}\label{eq:loss}
    \mathcal{L}_{cross}=-\sum_{r_{ui}\in\mathbf{R}}r_{ui}\text{log}(\sigma(\hat{y}_{ui}))+(1-r_{ui})\text{log}(1-\sigma(\hat{y}_{ui})),
\end{equation}
where $\mathbf{R}$ is the aforementioned user-item interaction matrix that records the ground-truth implicit feedback. With this loss, the optimal object turns to reconstruct the observed interactions in the training phase. 

In general, Eqn.($\ref{eq:loss}$) is effective in learning useful hash codes for users and items, and guarantees that the interacted user-item pairs have similar hash codes. However, the ranking among candidate items tends to be more important than their absolute scores, since recommender systems suggest items to users mainly according to their rankings. For implicit feedback data, the ranking relation can be obtained by sampling from the interaction matrix $\mathbf{R}$~\cite{rendle2012bpr}. Specifically, we assume that the observed interactions, which are more reflective of a user's preference, should be assigned with higher estimated values than the unobserved ones. Inspired by this, we introduce a ranking order reinforced loss function as:
\begin{equation}
    \mathcal{L}_{rank}=\sum_{(u,i,j)\in\mathcal{D}}\text{max}(0,-\sigma(\hat{y}_{ui})+\sigma(\hat{y}_{uj})+\alpha),
\end{equation}
where $\mathcal{D}=\{(u,i,j)|(u,i)\in\mathcal{R}^+,(u,j)\in\mathcal{R}^-\}$ denotes the triplet training data, $\mathcal{R}^+$ indicates the observed interactions, and $\mathcal{R}^-$ is the unobserved interactions; $\alpha$ is the margin parameter that helps to control the observed and unobserved interactions. By minimizing $\mathcal{L}_{rank}$, the interacted items will gather around the user $u$ more closely than items that are not interacted with $u$ in the Hamming space.

By combining the aforementioned factors, the overall loss function to be minimized is formulated as:
\begin{equation}
\begin{split}
    \mathcal{L} &= \mathcal{L}_{cross} + \lambda_1\mathcal{L}_{rank} + \lambda_2{\lVert\mathbf{E}\rVert}^2_2\\
    &=-\sum_{r_{ui}\in\mathbf{R}}r_{ui}\text{log}(\sigma(\hat{y}_{ui}))+(1-r_{ui})\text{log}(1-\sigma(\hat{y}_{ui}))\\
    &+\lambda_1\sum_{(u,i,j)\in\mathcal{D}}\text{max}(0,-\sigma(\hat{y}_{ui})+\sigma(\hat{y}_{uj})+\alpha) + \lambda_2{\lVert\mathbf{E}\rVert}^2_2,
\end{split}
\end{equation}
where $\lambda_1$ is the trade-off parameter to balance the proportion between the entropy and the ranking loss. Note that both the pair-wise ranking loss and the point-wise reconstruction loss are necessary for the model optimization; $\mathbf{E}\in\mathbb{R}^{(N+M)\times K}$ denotes the real-valued parameter matrix in the initial layer as trainable model parameters, and $\lambda_2$ controls the $L_2$ regularization strenghth to prevent overfitting. In addition, we adopt mini-batch Adam~\cite{kingma2014adam} to optimize our model and update the model parameters. Particularly, for a batch of randomly sampled triplets $(u,i,j)\in\mathcal{D}$, we obtain their final hash codes $\mathbf{h}_u^*$, $\mathbf{h}_i^*$, and $\mathbf{h}_j^*$, and then update model parameters by using the gradients of the loss function.

\subsection{\textbf{Comparison with HashGNN}}  We compare our hash code propagation layer with HashGNN~\cite{tan2020learning}, which adopts the embedding propagation in Euclidean space to learn hash codes. Hash\-GNN designs the graph convolution operation not on the hash code $\mathbf{h}_u^{(l)}$ (or $\mathbf{h}_i^{(l)}$) but on its real-valued parameter approximation $\mathbf{e}_u^{(l)}$ (or $\mathbf{e}_i^{(l)}$). There obviously exists a quantization deviation between them, \textit{i.e.}, $\mathbf{e}_u^{(l)}=\mathbf{h}_u^{(l)}+\mathbf{\epsilon}_u^{(l)}$. Hence, the propagation layer of HashGNN can be formulated as:
\begin{equation}
    \mathbf{e}_u^{(l+1)} = \text{tanh}\big(\mathbf{W}^{(l)}\cdot\text{mean}\big(\mathbf{h}_u^{(l)}+\mathbf{\epsilon}_u^{(l)}+\sum_{i\in \mathcal{N}_u}(\mathbf{h}_i^{(l)}+\mathbf{\epsilon}_i^{(l)})\big)\big),
\end{equation}
where the quantization deviations are accumulated with the aggregation. The nonlinear tanh function and weight matrix $\mathbf{W}^{(l)}\in \mathbb{R}^{K \times K}$ cause a new deviation between $\mathbf{e}_u^{(l+1)}$ and $\mathbf{h}_u^{(l+1)}$, which  will make the final hash codes unable to obtain the high-order Hamming similarity.
Compared with HashGNN, our proposed HS-GCN model implements the propagation of hash codes without introducing quantization deviations, effectively capturing the high-order Hamming similarity between hash codes. 




%% file: 3_experiment.tex
\section{Experimental Setup}
In this section, we first present the evaluation datasets, and then introduce our experimental settings, followed by elaboration of baseline methods.

\subsection{Datasets}
To justify the effectiveness of our proposed model, we conducted experiments on six publicly accessible datasets: MovieLens\footnote{https://grouplens.org/datasets/movielens/.}, Yelp\footnote{https://www.yelp.com/dataset.}, Amazon\footnote{http://jmcauley.ucsd.edu/data/amazon/.}, Gowalla, Pinterest, and Netflix\footnote{http://www.netflixprize.com/.}. Table~$\ref{tab:dataset}$ summarizes the detailed statistics of the evaluated datasets.

\noindent \textbf{MovieLens:} This is a widely used movie rating dataset, and we adopted the 1M version in our experiments. Similar to~\cite{he2016fast}, we transformed the rating scores into implicit feedback, so that the label of each user-item pair is either $1$ or $0$, indicating whether the user rated the movie. The other datasets are processed in a similar way.

\noindent \textbf{Yelp:} Yelp is a famous online review platform for business, such as restaurants, bars, and spas. We selected the dataset from the latest version and used a 20-core setting to provide a denser dataset.

\noindent \textbf{Amazon:} Amazon-review is a popular dataset for commodity recommendation~\cite{wang2017item}. We selected Amazon-book from the collection. Similarly, we used the $20$-core setting to ensure that each user and item have 
20 interactions at least.

\noindent \textbf{Gowalla}: This is the check-in dataset~\cite{liang2016modeling} obtained from Gowalla, where users share their locations by checking-in. To ensure the quality of the dataset, we used the 10-core setting, i.e., retaining users and items with at least ten interactions similar to~\cite{wang2019neural}.

\noindent \textbf{Pinterest}: This implicit feedback dataset is constructed by~\cite{geng2015learning} for evaluating image recommendation. We adopted its processed version shared by~\cite{he2017neural}.

\noindent \textbf{Netflix}: This is the large-scale movie rating dataset used in the Netflix challenge. We applied the 20-core setting to obtain a denser dataset.

For each dataset, we randomly split its implicit feedback into two parts: $70\%$ for training and the rest $30\%$ for testing. Moreover, $10\%$ interactions in the training set are randomly selected as the validation set for hyper-parameter tuning. For each observed user-item interaction in the training set, we treated it as a positive instance, and then adopted the negative sampling strategy to pair it with five negative items that the user did not interact with in the training set.

\begin{table}[]
    \centering
    \footnotesize
    \caption{Statistics of the datasets.}
    \vspace{-0.35cm}
    \begin{tabular}{|c|c|c|c|c|}
         \hline
         \rule{0pt}{10pt}Dataset & \#Users & \#Items & \#Interactions & Density  \\
         \hline
         \hline
         \rule{0pt}{10pt}MovieLens & 6,040 & 3,706 & 1,000,209 & 4.47\% \\
         \hline
         \rule{0pt}{10pt}Yelp & 40,500 & 58,755 & 2,024,283 & 0.09\% \\
         \hline
         \rule{0pt}{10pt}Amazon & 36,783 & 77,379 & 2,402,416 & 0.08\% \\
         \hline
         \rule{0pt}{10pt}Gowalla & 29,585 & 40,981 & 1,027,370 & 0.09\%\\
         \hline
         \rule{0pt}{10pt}Pinterest & 55,187 & 9,916 & 1,500,809 & 0.27\%\\
         \hline
         \rule{0pt}{10pt}Netflix & 429,584 & 17,764 & 99,884,887 & 1.31\%
         \\
         \hline
    \end{tabular}
    \vspace{-0.35cm}
    \label{tab:dataset}
\end{table}

\subsection{Experimental Settings}
\textbf{Evaluation Metric.} For each user in the testing set, we regarded all items with no interaction with her/him as negative ones. Then the recommendation method predicts the user's preference scores (\emph{e.g.}, Hamming similarity) over all the items, except the ones used in the training set. To measure the performance of top-$K$ recommendation and preference ranking, we adopted HR@$K$ (Hit Ratio)~\cite{he2017neural} and NDCG@$K$ (Normalized Discounted Cumulative Gain) as the evaluation metrics. Whereinto, NDCG@$K$ are formally defined as:
\begin{equation*}
    \text{NDCG}@K = \frac{\text{DCG}_K}{\text{IDCG}_K},~\text{and}~
    \text{DCG}_K = \sum_{i=1}^K\frac{2^{r_{ui}}-1}{\text{log}_2(1+i)},
\end{equation*}
where IDCG is the ideal DCG, and $r_{ui}$ denotes the interaction status of the $i$-th recommended item. 
By default, we set $K=50$ and $100$. We reported the average metrics for all users in the testing set.

\noindent \textbf{Model Implementation.} We implemented our HS-GCN model via the development tool Pytorch\footnote{https://pytorch.org/.} and Pytorch Geometric\footnote{https://pytorch-geometric.readthedocs.io/en/latest/.}. We set the depth of HS-GCN $L$ as two to model the second-order Hamming similarity. 
The hash code size is set to be $64$. We optimized our model by Adam optimizer with a batch size of 3,000. Besides, we utilized the popular approach of Xavier~\cite{glorot2010understanding} to initialize all the parameters in our model. And we applied the grid search for tuning the hyper-parameters based on the results from the validation set: the learning rate is tuned amongst \{1e-4, 3e-4, 1e-3, 3e-3\}, and finally set to be 3e-4; the trade-off coefficient $\lambda_1$ is tuned from 0.1 to 1 with step size of 0.1, and finally set to be 0.1; the coefficient $\lambda_2$ of $L_2$ normalization is searched within \{1e-8, 1e-7, $\cdots$, 1\}, and 1e-7 is the optimal value. The margin parameter $\alpha$ is fixed to $0.2$. Moreover, early stopping is adopted if HR@$K$ on the validation data does not increase for 10 successive epochs.

\subsection{Baseline Comparison}
To demonstrate the effectiveness, we compared the performance of our proposed method with a series of state-of-the-art hash learning based models (\textit{i.e.}, DHN, Hash\_ste, HashNet, and HashGNN). Beyond these methods, we also introduce several real-valued recommendation models (\textit{i.e.}, MF, BiNE, PinSage, and GraphSAGE) to justify that HS-GCN could achieve comparable results.  
\begin{itemize}[leftmargin=*]
\item \textbf{DCF}~\cite{zhang2016discrete}: It is the first method to directly optimize hash codes based on the rating matrix. In our experiments, we adapted DCF for addressing the interaction matrix.
\item \textbf{DFM}~\cite{liu2018discrete}: This is a discretely parameterized factorization machine for rating prediction. We adapted it to be trained on implicit data.
\item \textbf{DHN}~\cite{zhu2016deep}: This is a popular deep hashing method for similar image recommendation. It learns high-quality hash codes by controlling the quantization deviation. We adapted it for user-item recommendation by replacing the original AlexNet~\cite{krizhevsky2017imagenet} framework with Graph Convolution Networks~(GCNs).
\item \textbf{Hash\_ste}: This is an effective end-to-end hash learning method based on straight through estimator~\cite{bengio2013estimating}, which enables the optimization of discrete variables by directly copying approximated gradients of them.  
\item \textbf{HashNet}~\cite{cao2017hashnet}: This is a state-of-the-art deep hashing method, which devises a continuous and differentiable function to approximate the sign function. Similarly, we used GCN to replace the AlexNet in HashNet.
\item \textbf{GCNH}~\cite{zhou2018graph}: It is a graph-based hashing method, which introduces a novel and efficient asymmetric graph convolution network, followed by a binarization step to learn similarity-preserving hash codes. We adapted it for recommendation via the user and item ID information as its input feature.
\item  \textbf{DGCN-BinCF}~\cite{wang2019binarized}: It is a graph-based hashing method. This method relaxes the binary constraint and makes continuous optimization possible to distill the ranking information from GCN into hash codes.
\item \textbf{HashGNN}~\cite{tan2020learning}: It is a graph-based hashing model, where GNN is utilized towards high quality hash codes. 
\item \textbf{MF}~\cite{mnih2008probabilistic}: Matrix Factorization is the most common embedding model for recommendation, which only exploits the user-item direct interactions for recommendation.
\item \textbf{BiNE}~\cite{gao2018bine}: This embedding model adopts biased random walks for representation learning based on the bipartite graph. It applies an optimization framework for both explicit ratings and implicit feedback. 
\item \textbf{PinSage}~\cite{ying2018graph}: PinSage is a graph embedding model. It combines efficient random walks and graph convolutions on the item-item graph, which incorporates both the graph structure and the node feature information. Hereon, we applied it on the user-item interaction graph.
\item \textbf{GraphSage}~\cite{hamilton2017inductive}: As a famous graph embedding model, it is a general inductive framework that learns embeddings by sampling and aggregating features from a node's local neighborhood. 
\end{itemize}

\noindent \textbf{Baseline Implementation.} For fair comparison, all baseline methods are implemented in Pytorch and Pytorch Geometric, except the publicly available implementation of DCF\footnote{https://github.com/hanwangzhang/Discrete-Collaborative-Filtering.}, DFM\footnote{https://github.com/hanliu95/DFM.}, GCNH\footnote{https://github.com/zxJohnFly/GCN.}, and BiNE\footnote{https://github.com/clhchtcjj/BiNE.}. Without specification, the default size of embedding or hash code is $64$. For all graph-based baselines, we adopted a two-layer GCN initialized from Xavier, and used the Adam optimizer with a well-chosen mini-batch size for model optimization. The learning rate is tuned amongst \{1e-4, 3e-4, 1e-3, 3e-3\}. For HashGNN, we set the architecture of the graph layer, and the hyper-parameters, following the original paper~\cite{tan2020learning}. For HashNet, we initialized the scale parameter $\beta$ for tanh$(\beta\mathbf{x})$ as 1, and exponentially increased it after constant epochs as suggested in~\cite{cao2017hashnet}. For DGCN-BinCF, PinSage, and GraphSage, we implemented them according to the default architectures in the original papers~\cite{wang2019binarized,gao2018bine, ying2018graph,hamilton2017inductive}, and tuned their hyper-parameters based upon the model performance on validation set.

%% file: 4_results.tex
\section{Experimental Results}
In order to validate the effectiveness of our proposed method, we conducted overall experiments to compare our proposed HS-GCN model with baseline methods in performance, efficiency, and sparsity issue. Moreover, in order to evaluate the effectiveness of components in HS-GCN, we performed ablation experiments on the key components including: propagation layers, ranking loss, and dropout technique.

\begin{table*}[t]\centering
\footnotesize
\caption{Overall performance comparison on six datasets. \% Improv. and p-value denote the relative improvements (\%) and t-test results of HS-GCN compared with HashGNN, respectively.}
\vspace{-0.35cm}
\label{tab:overall}
\setlength{\tabcolsep}{1.4mm}{
\begin{tabular}{|l|cc|cc|cc|cc|cc|cc|}
     \hline
     \rule{0pt}{10pt}
     \multirow{2}*{Methods} & \multicolumn{2}{c|}{MovieLens} & \multicolumn{2}{c|}{Yelp} & \multicolumn{2}{c|}{Amazon} & \multicolumn{2}{c|}{Gowalla} & \multicolumn{2}{c|}{Pinterest} & \multicolumn{2}{c|}{Netflix}\\
     \rule{0pt}{10pt}
     {} & HR@50 & HR@100 & HR@50 & HR@100 & HR@50 & HR@100 & HR@50 & HR@100 & HR@50 & HR@100 & HR@50 & HR@100\\
     \hline
     \hline
     \rule{0pt}{10pt}DCF & 0.0469 & 0.0910 & 0.0143 & 0.0281 & 0.0145 & 0.0288 & 0.1376 & 0.1471 & 0.0723 & 0.0840 & 0.0667 & 0.1143\\
     \rule{0pt}{10pt}DFM & 0.0549 & 0.1068 & 0.0163 & 0.0310 & 0.0171 & 0.0336 & 0.1753 & 0.1820 & 0.0849 & 0.0931 & 0.0819 & 0.1330\\
     \rule{0pt}{10pt}DHN & 0.0782 & 0.1393 & 0.0228 & 0.0430 & 0.0238 & 0.0448 & 0.2281 & 0.2415 & 0.1200 & 0.1328 & 0.1084 & 0.1813\\
     \rule{0pt}{10pt}Hash\_ste & 0.0955 & 0.1684 & 0.0349 & 0.0620 & 0.0327 & 0.0557 & 0.3268 & 0.3606 & 0.1125 & 0.2032 & 0.1211 & 0.2007\\ 
     \rule{0pt}{10pt}HashNet & 0.1359 & 0.2232 & 0.0364 & 0.0651 & 0.0353 & 0.0599 & 0.3370 & 0.3830 & 0.1539 & 0.2730 & 0.1463 & 0.2337\\
     \rule{0pt}{10pt}GCNH & 0.1552 & 0.2488 & 0.0389 & 0.0714 & 0.0394 & 0.0636 & 0.3979 & 0.4209 & 0.1847 & 0.3129 & 0.1515 & 0.2402\\
     \rule{0pt}{10pt}DGCN-BinCF & 0.1554 & 0.2546 & 0.0358 & 0.0666 & 0.0373 & 0.0623 & 0.3420 & 0.3853 & 0.1834 & 0.3112 & 0.1744 & 0.2883\\
     \rule{0pt}{10pt}\underline{HashGNN} & \underline{0.1598} & \underline{0.2557} & \underline{0.0447} & \underline{0.0774} &  \underline{0.0363} & \underline{0.0611} & \underline{0.3932} & \underline{0.4329} & \underline{0.1947} & \underline{0.3214} & \underline{0.2058} & \underline{0.3185}\\
     \rule{0pt}{10pt}\rule{0pt}{10pt}\textbf{HS-GCN} & \textbf{0.2052} & \rule{0pt}{10pt}\textbf{0.3169} & \textbf{0.0497} & \textbf{0.0883} & \textbf{0.0523} & \textbf{0.0885} & \textbf{0.4084} & \textbf{0.4480} & \textbf{0.2066} & \textbf{0.3322} & \textbf{0.2192} & \textbf{0.3432}\\
     \hline
     \rule{0pt}{10pt}MF & 0.1332 & 0.2042 & 0.0319 & 0.0513 & 0.0349 & 0.0566 & 0.3360 & 0.3675 & 0.1400 & 0.2472 & 0.1981 & 0.3042\\
     \rule{0pt}{10pt}BiNE & 0.1308 & 0.1868 & 0.0345 & 0.0546 & 0.0335 & 0.0560 & 0.3603 & 0.3812 & 0.1526 & 0.2582 & 0.2097 & 0.3346 \\
     \rule{0pt}{10pt}PinSage & 0.1587 & 0.2501 & 0.0503 & 0.0812 & 0.0402 & 0.0656 & 0.3791 & 0.4131 & 0.1837 & 0.3008 & 0.2279 & 0.3585\\
     \rule{0pt}{10pt}GraphSage & 0.2132 & 0.3326 & 0.0581 & 0.0942 & 0.0408 & 0.0672 & 0.3894 & 0.4305 & 0.1919 & 0.3162 & 0.2334 & 0.3616\\
     \hline
     \rule{0pt}{10pt}\% Improv. & 28.41\% & 23.93\% & 11.19\% & 14.08\% & 44.08\% & 44.84\% & 3.87\% & 3.49\% & 6.11\% & 3.36\% & 6.51\% & 7.76\%\\
     \rule{0pt}{10pt}p-value & 2.77e-4 & 7.01e-4 & 5.62e-3 & 1.95e-4 & 1.04e-3 & 1.27e-4 & 1.81e-4 & 1.39e-4 & 1.37e-3 & 5.08e-4 & 1.35e-4 & 2.04e-4\\
     \hline
     \hline
     \rule{0pt}{10pt}
     \multirow{2}*{Methods} & \multicolumn{2}{c|}{MovieLens} & \multicolumn{2}{c|}{Yelp} & \multicolumn{2}{c|}{Amazon} & \multicolumn{2}{c|}{Gowalla} & \multicolumn{2}{c|}{Pinterest} & \multicolumn{2}{c|}{Netflix}\\
     \rule{0pt}{10pt}
     {} & N@50 & N@100 & N@50 & N@100 & N@50 & N@100 & N@50 & N@100 & N@50 & N@100 & N@50 & N@100\\
     \hline
     \hline
     \rule{0pt}{10pt}DCF & 0.0641 & 0.0785 & 0.0097 & 0.0137 & 0.0113 & 0.0173 & 0.2066 & 0.2599 & 0.0756 & 0.1109 & 0.0862 & 0.1014\\
     \rule{0pt}{10pt}DFM & 0.0767 & 0.0937 & 0.0105 & 0.0165 & 0.0131 & 0.0187 & 0.2616 & 0.3041 & 0.0912 & 0.1242 & 0.1019 & 0.1294\\
     \rule{0pt}{10pt}DHN & 0.1038 & 0.1251 & 0.0149 & 0.0226 & 0.0176 & 0.0264 & 0.3440 & 0.4080 & 0.1212 & 0.1710 & 0.1373 & 0.1660\\
     \rule{0pt}{10pt}Hash\_ste & 0.1305 & 0.1608 & 0.0225 & 0.0322 & 0.0263 & 0.0354 & 0.4692 & 0.5114 & 0.1248 & 0.1748 & 0.1515 & 0.1817\\ 
     \rule{0pt}{10pt}HashNet & 0.1940 & 0.2233 & 0.0232 & 0.0335 & 0.0281 & 0.0378 & 0.4825 & 0.5245 & 0.1441 & 0.1877 & 0.1710 & 0.1900\\
     \rule{0pt}{10pt}GCNH & 0.1443 & 0.2283 & 0.0294 & 0.0450 & 0.0319 & 0.0408 & 0.4957 & 0.5490 & 0.1616 & 0.2105 & 0.1723 & 0.2049\\
     \rule{0pt}{10pt}DGCN-BinCF & 0.2298 & 0.2688 & 0.0268 & 0.0345 & 0.0302 & 0.0390 & 0.4646 & 0.5234 & 0.1436 & 0.2072 & 0.2230 & 0.2745\\
     \rule{0pt}{10pt}\underline{HashGNN} & \underline{0.2335} & \underline{0.2630} & \underline{0.0308} & \underline{0.0430} & \underline{0.0273} & \underline{0.0374} & \underline{0.5155} & \underline{0.5512} & \underline{0.1693} & \underline{0.2198} & \underline{0.3056} & \underline{0.3502}\\
     \rule{0pt}{10pt}\textbf{HS-GCN} & \textbf{0.3081} & \rule{0pt}{10pt}\textbf{0.3376} & \textbf{0.0336} & \textbf{0.0483} & \textbf{0.0436} & \textbf{0.0585} & \textbf{0.5351} & \textbf{0.5696} & \textbf{0.1786} & \textbf{0.2325} & \textbf{0.3883} & \textbf{0.4501}\\
     \hline
     \rule{0pt}{10pt}MF & 0.2023 & 0.2216 & 0.0240 & 0.0312 & 0.0303 & 0.0392 & 0.4977 & 0.5221 & 0.1029 & 0.1514 & 0.2469 & 0.2883\\
     \rule{0pt}{10pt}BiNE & 0.1370 & 0.1765 & 0.0231 & 0.0320 & 0.0278 & 0.0370 & 0.5104 & 0.5346 & 0.1169 & 0.1858 & 0.2584 & 0.2999\\
     \rule{0pt}{10pt}PinSage & 0.2316 & 0.2517 & 0.0380 & 0.0496 & 0.0337 & 0.0438 & 0.5136 & 0.5447 & 0.1536 & 0.2063 & 0.2923 & 0.3261\\
     \rule{0pt}{10pt}GraphSage & 0.3195 & 0.3490 & 0.0434 & 0.0570 & 0.0349 & 0.0457 & 0.5368 & 0.5581 & 0.1643 & 0.2124 & 0.3703 & 0.4114\\
     \hline
     \rule{0pt}{10pt}\% Improv. & 31.95\% & 28.37\% & 9.09\% & 12.33\% & 59.71\% & 56.42\% & 3.80\% & 3.34\% & 5.49\% & 5.78\% & 27.1\% & 28.5\%\\
     \rule{0pt}{10pt}p-value & 3.28e-3 & 2.87e-3 & 1.41e-2 & 8.47e-4 & 3.51e-4 & 1.44e-4 & 2.21e-3 & 4.09e-4 & 4.11e-3 & 1.69e-3 & 4.15e-3 & 3.72e-3\\
     \hline
\end{tabular}}
\vspace{-0.35cm}
\end{table*}

\subsection{Overall Experiments}
We started by comparing the performance of all the methods, and then analyzed the efficiency of these methods. Finally, we explored the effectiveness of modeling high-order similarity under the sparse settings.
\subsubsection{Performance Comparison}
The results of our method and baselines over the experimented datasets are presented in Table $\ref{tab:overall}$. Besides, we reported the improvements and statistical significance test in Table $\ref{tab:overall}$, which are calculated between our proposed method and the strongest hashing baseline (highlighted with underline). Observing the results shown in Table $\ref{tab:overall}$ from top to bottom, we obtained the following key findings: 

\begin{itemize}[leftmargin=*]
    \item DCF and DFM underperform on the implicit feedback datasets. This reflects the limitation of hashing methods that are designed on the basis of explicit feedback. DHN obtains poor performance on all the datasets. This indicates that the two-stage hashing method is insufficient to capture first-order similarities between users and items in the Hamming space, which is caused by the quantization deviation. Hash\_ste consistently exceeds DHN across all cases, demonstrating the advantage of directly learning to hash strategy. However, Hash\_ste fails to completely solve the quantization deviation between the straight through estimator and the sign function. 
    \item Compared with DHN and Hash\_ste, HashNet effectively eliminates the quantization deviation, and the better performance verifies that HashNet captures the first-order Hamming similarity more accurately. However, none of these methods explicitly models the high-order Hamming similarity in the hash code learning process, which leads to suboptimal results.
    \item GNN-based hashing methods significantly outperform the hashing methods based on deep learning. It makes sense since GNN improves the quality of user and item continuous representations prior to the binarization step, which indirectly enriches the final hash codes. Within GNN-based methods, HashGNN is superior since it is a real sense of end-to-end hash learning method, due to its ability of solving the back-propagation issue of sign function by gradient copy. In contrast, DGCN-BinCF and GCNH both use sign function as an extra step for binarization, inevitably resulting in quantization loss.
    \item It is clear that HS-GCN yields the best performance among hashing methods. In particular, the average improvements of HS-GCN over the strongest hashing baseline HashGNN \emph{w.r.t.} HR@$K$ are 26.17$\%$, 12.64$\%$, 44.46$\%$, 3.68$\%$, 4.74$\%$, and 7.14$\%$, and \emph{w.r.t.} NDCG@$K$ are 30.16$\%$, 10.71$\%$, 58.07$\%$, 3.57$\%$, 5.64$\%$, and 27.8$\%$ in MovieLens, Yelp, Amazon, Gowalla, Pinterest, and Netflix, respectively. The reason is that HS-GCN is capable of capturing the high-order similarity of hash codes by directly constructing GCN in the Hamming space, while HashGNN only uses the first-order similarity to guide the hash learning.
    Additionally, we conducted the cross validation t-test, and p-value $< 0.05$ indicates that the improvements of HS-GCN over the strongest hashing baseline are statistically significant.
    \item HS-GCN significantly outperforms the embedding based baselines MF and BiNE. Moreover, compared with the graph embedding models GraphSage and PinSage, our model yields comparable performance. GraphSage performs better than most hash-based methods, since it utilizes GCN to learn real-valued embeddings for users and items, which have better representative ability than binary codes. It is worth mentioning that our performance on Amazon is superior than the best graph embedding model GraphSage. This also verifies the importance of capturing the high-order similarity for hash learning. 
\end{itemize}


\begin{figure*}[tbp]
\centering
\begin{minipage}[t]{0.33\linewidth}
\centering
\includegraphics[width=\linewidth]{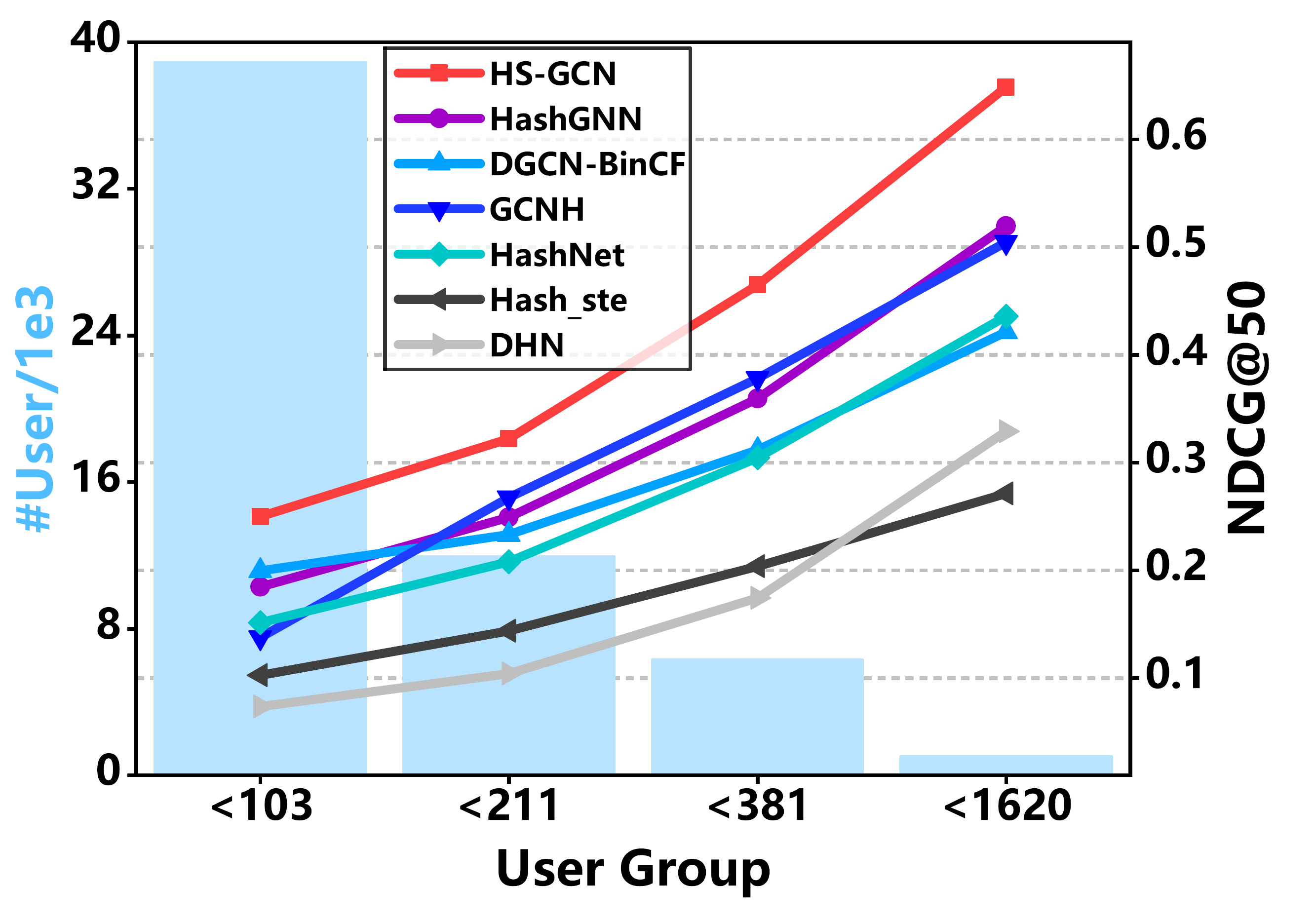}
\centerline{(a)~MovieLens}
\end{minipage}%
\begin{minipage}[t]{0.33\linewidth}
\centering
\includegraphics[width=\linewidth]{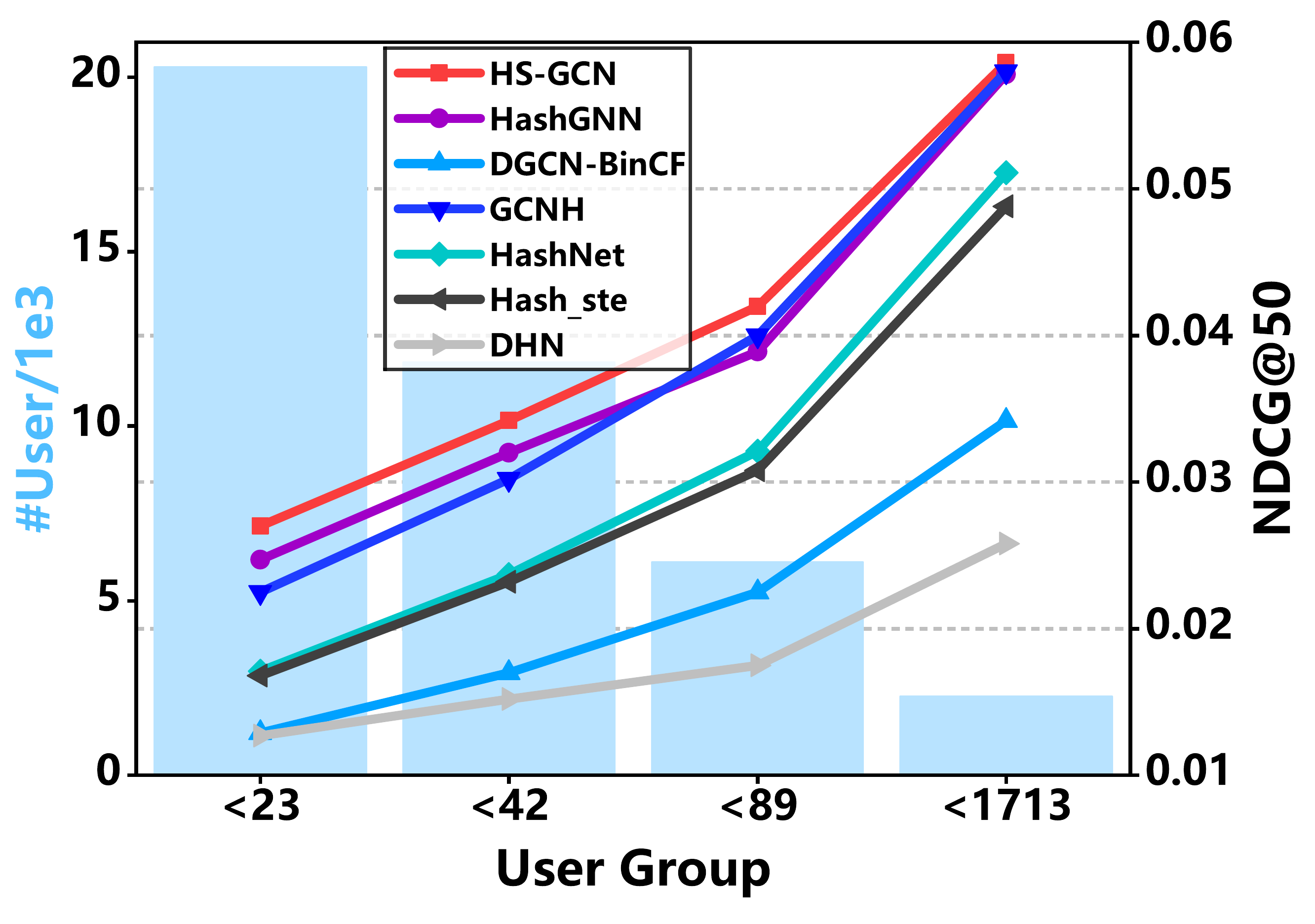}
\centerline{(b)~Yelp}
\end{minipage}%
\begin{minipage}[t]{0.33\linewidth}
\centering
\includegraphics[width=\linewidth]{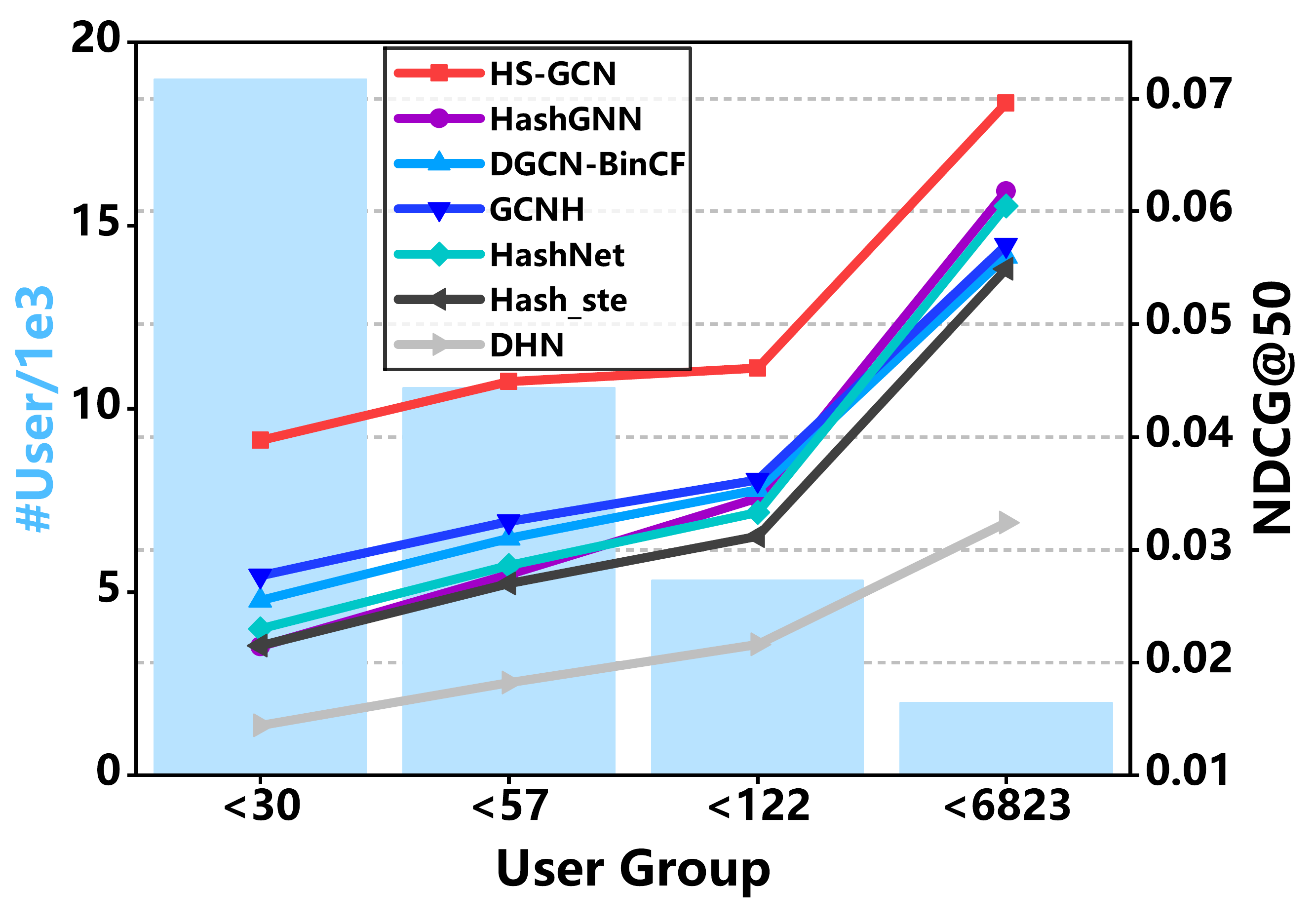}
\centerline{(c)~Amazon}
\end{minipage}%
\centering
\vspace{-0.2cm}
\caption{Performance comparison over the sparsity distribution of user groups on three different datasets. Wherein, the background
histograms indicate the number of users involved in each group, and the lines demonstrate the performance \emph{w.r.t.} NDCG@50.}
\label{fig:sparsity}
\vspace{-0.5cm}
\end{figure*}

\subsubsection{Efficiency Comparison}
In this section, we analyzed the efficiency of HS-GCN compared with baselines. In particular, we focused on both the training and the testing efficiency of these models, and recorded their running time. For fair comparison, all the models are trained on Ubuntu 16.04.5 with NVIDIA TITAN Xp, 12GB frame buffer and Python3.7 until the convergence, and tested on Windows 10 with Intel Core i7, 16GB RAM and Python3.7. During the training process, the hyper-parameters (\emph{e.g.}, the training epoch, the batch size, and the learning rate) of all models follow the optimal settings. Though we adopted linear scan as the testing technique to recommend items for all methods, it is worth mentioning some techniques like multi-level indexing~\cite{norouzi2012fast} can further accelerate the testing of hashing methods. 
Table~\ref{tab:efficiency} summarizes the results on different-scale Amazon dataset, where ``1/8Amazon'' is constructed by randomly selecting one in eight interactions from the complete dataset, and the rest can be done in the same manner. There are similar efficiency comparisons on other datasets, and thus we omitted them for space saving. Regarding the experiments, we found that:
\begin{itemize}[leftmargin=*]
    \item Our model costs much less training time than other methods. Compared to MF, HS-GCN and other graph-based models are more efficient in training since they quickly converge in experiments. This is reasonable since indirectly connected users and items are involved through the graph structure when optimizing the interaction pairs in mini-batch. Compared to other graph-based models, HS-GCN performs more efficiently during the training process, since it has the most concise model size that is the same as MF. Specifically, HS-GCN introduces no additional parameters except for the initial real-valued matrix $\mathbf{E}$, while other models introduce the trainable weight matrix as transfer parameters in each propagation layer.
    \item Similar with most hashing models, HS-GCN achieves significant speedup compared with continuous recommendation models regarding the testing time. By jointly analyzing Table $\ref{tab:overall}$ and Table~$\ref{tab:efficiency}$, we can see that HS-GCN not only can achieve competitive performance compared to state-of-the-art continuous models, but also is time-efficient. This verifies that HS-GCN is an operable solution for large-scale Web services to substantially reduce the computation cost of their recommendation systems~\cite{liu2014personalized}.
    \item Across the different-scale datasets, the training efficiency ratio of HS-GCN is stable around 2.4 times  based on HashGNN, and the testing efficiency ratio is stable around 4.9 times based on GraphSage. From Table~\ref{tab:efficiency}, we can observe that both training and testing efficiency ratios of our method show upward trends with the scale of the dataset gradually increasing.
\end{itemize}

\begin{table}[t]\scriptsize
\caption{Training/Testing time on different-scale Amazon dataset with $64$ bits. The results are reported in hours and seconds, respectively. Training/Testing efficiency ratio is computed between HS-GCN and HashGNN/GraphSage, which is the best graph-based hashing/embedding baseline.}
\vspace{-0.35cm}
    \centering
    \setlength{\tabcolsep}{1.3mm}{
    \begin{tabular}{|l|r|r|r|r|r|r|r|r|}
         \hline
         \rule{0pt}{10pt}
         \multirow{2}*{Methods} & \multicolumn{2}{c|}{1/8Amazon} & \multicolumn{2}{c|}{1/4Amazon} & \multicolumn{2}{c|}{1/2Amazon} & \multicolumn{2}{c|}{Amazon}\\
         \cline{2-9}
         \rule{0pt}{10pt}
         {} & Train & Test & Train & Test & Train & Test & Train & Test\\
         \hline
         \hline
         \rule{0pt}{10pt}
         \scriptsize{HS-GCN} & 1.3 & 3.0 & 2.4 & 6.1 & 4.6 & 13.3 & 9.1 & 24.8 \\
         \rule{0pt}{10pt}
         \scriptsize{DHN} & 1.4 & 3.0 & 2.7 & 6.1 & 5.4 & 13.2 & 11.2 & 24.9 \\
         \rule{0pt}{10pt}
         Hash\_ste & 1.5 & 3.0 & 3.0 & 6.0 & 6.1 & 13.3 & 12.6 & 24.7  \\
         \rule{0pt}{10pt}
         HashNet & 2.1 & 3.0 & 4.3 & 6.2 & 9.0 & 13.1 & 18.3 & 24.8 \\
         \rule{0pt}{10pt}
         GCNH & 1.9 & 3.1 & 3.6 & 6.1 & 7.2 & 13.3 & 14.8 & 25.0 \\
         \rule{0pt}{10pt}
         \scriptsize{DGCN-BinCF} & 1.6 & 3.1 & 3.2 & 6.1 & 6.2 & 13.3 & 12.6 & 24.8 \\
         \rule{0pt}{10pt}
         HashGNN & 2.7 & 3.0 & 5.6 & 6.1 & 11.1 & 13.2 & 23.3 & 24.9 \\
         \hline
         \rule{0pt}{10pt}
         MF & 2.3 & 14.1 & 4.3 & 30.5 & 9.1 & 66.2 & 18.9 & 127.3  \\
         \rule{0pt}{10pt}
         BiNE & 1.4 & 14.2 & 2.6 & 30.2 & 5.4 & 66.2 & 10.9 & 127.0 \\
         \rule{0pt}{10pt}
         PinSage & 1.6 & 14.1 & 3.2 & 30.4 & 6.5 & 66.0 & 13.1 & 127.2 \\
         \rule{0pt}{10pt}
         GraphSage & 1.5 & 14.1 & 2.9 & 30.1 & 6.0 & 65.9 & 12.1 & 127.0 \\
         \hline
         \rule{0pt}{10pt}
         \scriptsize{Efficiency Ratio} & 2.1 & 4.7 & 2.3 & 4.9 & 2.4 & 5.0 & 2.6 & 5.1  \\
         \hline
    \end{tabular}
    }
    \label{tab:efficiency}
    \vspace{-0.35cm}
\end{table}

\subsubsection{Performance Comparison w.r.t. Interaction Sparsity Levels}
The sparsity issue usually limits the expressiveness of recommender systems, since inactive users lack sufficient interactions to generate high-quality representations. This especially impedes the learning to hash methods only based on the first-order Hamming similarity. In this section, we attempted to answer whether exploring the high-order Hamming similarity is useful to alleviate this issue.

Towards this end, we first divided users into different sparsity-level groups. In particular, according to the interaction number per user in the training set, we divided the testing set into four groups, each of which has the same interaction sum. Taking Yelp dataset as an example, the interaction numbers per user are less than 23, 42, 89, and 1,713, respectively. Figure $\ref{fig:sparsity}$ shows the experiment results \emph{w.r.t.} NDCG@50 on different user groups in MovieLens, Yelp, and Amazon. There is a similar performance trend \emph{w.r.t.} HR@50, and thus we omitted this part for space saving. Regarding the results, we found that:

\begin{itemize}[leftmargin=*]
    \item HS-GCN consistently outperforms all other hashing baselines on all user groups. This demonstrates that exploiting the high-order Hamming similarity can improve the hash code learning for both active and inactive users.
    \item After analyzing Figures $\ref{fig:sparsity}$, we observed that HS-GCN achieves larger improvements in the first two groups (\emph{e.g.}, $9.31\%$ and $6.88\%$ over the best baseline separately for $<23$ and $<42$ in Yelp) than that of the others (\emph{e.g.}, $0.52\%$ for $<$1,713 Yelp group). It verifies that capturing the high-order Hamming similarity is especially beneficial to the inactive users, since their hash codes are learned from more sufficient similar information besides first-order similarities. Hence, exploiting the high-order similarity is promising to solve the sparsity issue in hashing-based recommender systems.
\end{itemize}

\subsection{Ablation Experiments}
In this section, we studied the effect of key components in our proposed model, including propagation layers, ranking loss, and dropout. As the hash code propagation layer plays an important role in our model, we started by investigating the influence of layer numbers on the performance. Then we analyzed the ranking reinforced loss in the model optimization. Moreover, we jointly analyzed the affects of node dropout and bit dropout ratios.
\begin{table}[t]\centering
\footnotesize
\caption{Effect of hash codes propagation layer numbers ($L$).}
\vspace{-0.35cm}
\label{tab:layer}
\setlength{\tabcolsep}{1.5mm}{
\begin{tabular}{|l|cc|cc|cc|}
     \hline
     \rule{0pt}{10pt}\multirow{2}*{Methods} & \multicolumn{2}{c|}{MovieLens} & \multicolumn{2}{c|}{Yelp} & \multicolumn{2}{c|}{Amazon} \\
      \rule{0pt}{10pt}& \footnotesize HR@50 & \footnotesize N@50 & \footnotesize HR@50 & \footnotesize N@50 & \footnotesize HR@50 & \footnotesize N@50 \\
     \hline
     \hline
     \rule{0pt}{10pt}HS-GCN-1 & 0.1822 & 0.2774 & 0.0452 & 0.0316 & 0.0407 & 0.0340\\ 
     \rule{0pt}{10pt}HS-GCN-2 & 0.2052 & 0.3081 & 0.0497 & 0.0336 & 0.0523 & 0.0436\\
     \rule{0pt}{10pt}HS-GCN-3 & 0.1840 & 0.2798 & 0.0482 & 0.0329 & 0.0500 & 0.0421\\
     \hline
\end{tabular}}
\end{table}

\subsubsection{Effect of Layer Numbers}
To investigate the optimal number of hash code propagation layers, we varied the model depth. Particularly, we searched the layer numbers within $\{1, 2, 3\}$. Table $\ref{tab:layer}$ summarizes the experimental results, where HS-GCN-2 denotes the model with two propagation layers, and similar notations for the others. Jointly analyzing Table $\ref{tab:overall}$ and Table $\ref{tab:layer}$, we have the following observations:
\begin{itemize}[leftmargin=*]
\item Obviously, HS-GCN-2 and HS-GCN-3 consistently outperform HS-GCN-1 across all datasets. The reason is that HS-GCN-1 only considers the first-order Hamming similarity, while HS-GCN-2 and HS-GCN-3 both utilize the high-order Hamming similarity.
\item HS-GCN-2 achieves the best performance. When further stacking the propagation layer, we found that HS-GCN-3 leads to overfitting on all the datasets. The optimal layer number of our model is consistent with the prior HashGNN~\cite{tan2020learning}. Therefore, the second-order similarity is sufficient for hash learning.
\item When varying the number of propagation layers, HS-GCN is consistently superior to the learning to hash baselines across three datasets. This verifies that directly capturing the high-order Hamming similarity can facilitate the quality of hash codes.
\end{itemize}


\begin{table}[t]\centering
\footnotesize
\caption{Performance comparison of the models with and without different losses.}
\vspace{-0.29cm}
\label{tab:rank_loss}
\setlength{\tabcolsep}{1.0mm}{
\begin{tabular}{|l|cc|cc|cc|}
     \hline
     \rule{0pt}{10pt}
     \multirow{2}*{Methods} & \multicolumn{2}{c|}{MovieLens} & \multicolumn{2}{c|}{Yelp} & \multicolumn{2}{c|}{Amazon} \\
     \rule{0pt}{10pt}
      & \footnotesize HR@50 & \footnotesize N@50 & \footnotesize HR@50 & \footnotesize N@50 & \footnotesize HR@50 & \footnotesize N@50 \\
     \hline
     \hline
     \rule{0pt}{10pt}HS-GCN & 0.2052 & 0.3081 & 0.0497 & 0.0336 & 0.0523 & 0.0436\\
     \rule{0pt}{10pt}W/o Initial Rank & 0.1773 & 0.2747 & 0.0468 & 0.0313 & 0.0509 & 0.0428\\
     \rule{0pt}{10pt}W/o Final Rank & 0.1769 & 0.2566 & 0.0418 & 0.0281 & 0.0460 & 0.0385\\
     \rule{0pt}{10pt}W/o CR-Entropy & 0.1564 & 0.2426 & 0.0156 & 0.0100 & 0.0289 & 0.0239 \\
     \hline
\end{tabular}}
\vspace{-0.3cm}
\end{table}

\subsubsection{Effect of the Ranking Loss}
For capturing the relative item ranking in hash learning, we introduced the ranking reinforced loss in model optimization. During the experiments, we found that the ranking loss can contribute to both the initial layer and the prediction layer, where the ranking losses are computed by the initial state $\{\mathbf{h}_u, \mathbf{h}_i, \mathbf{h}_j\}$ and the final hash codes $\{\mathbf{h}_u^{*}, \mathbf{h}_i^{*}, \mathbf{h}_j^{*}\}$, respectively. To study the influence of these two types of ranking losses, we conducted ablation study to validate how each of them contributes to the overall performance of our model.

Table~\ref{tab:rank_loss} records the results of the ablation study on three datasets, where HS-GCN is our intact model with two types of ranking losses, which are computed upon triplets of the initial state $\{\mathbf{h}_u, \mathbf{h}_i, \mathbf{h}_j\}$ and the final hash codes $\{\mathbf{h}_u^{*}, \mathbf{h}_i^{*}, \mathbf{h}_j^{*}\}$, respectively. W/o Initial Rank denotes the ablation variant of HS-GCN via removing the ranking loss on the triplets of initial hash codes, while W/o Final Rank is the variant via removing the ranking loss on the final hash codes.
From the table, we can observe that HS-GCN performs worse when ignoring the ranking loss for prediction. This demonstrates the capacity of ranking loss for more effective hashing. Moreover, we found that the ranking loss can improve the model performance more significantly when acting on the initial layer, since it injects the order information into the initial state. This also demonstrates that initialization enhances the quality of final hash codes, which is consistent with the
findings of prior efforts~\cite{zhang2016discrete,lian2017discrete}.

Following the operations in HashGNN~\cite{tan2020learning}, we introduce the cross-entropy loss to supervise the interactions between users and items they consumed before. We suggest that the cross-entropy loss maximization provides the signal to preserve the bipartite graph structure, which is the base of graph neural networks. Since this is not originally presented by us, we did not claim this design in our paper.
To evaluate the effectiveness of each design, we do ablation study on the several datasets. In particular, we  removed the binary cross-entropy loss from the final loss function, denoted as W/o CR-Entropy in Table~\ref{tab:rank_loss}, and observed that the variant is suboptimal compared with the intact one. 
The possible reason is that the binary cross-entropy loss is capable of supervising the hash codes to reconstruct the interactions between users and items, and thus optimizes their hash coding. 

\begin{figure}[tbp]
\centering
\begin{minipage}[t]{0.33\linewidth}
\centering
\includegraphics[width=\linewidth]{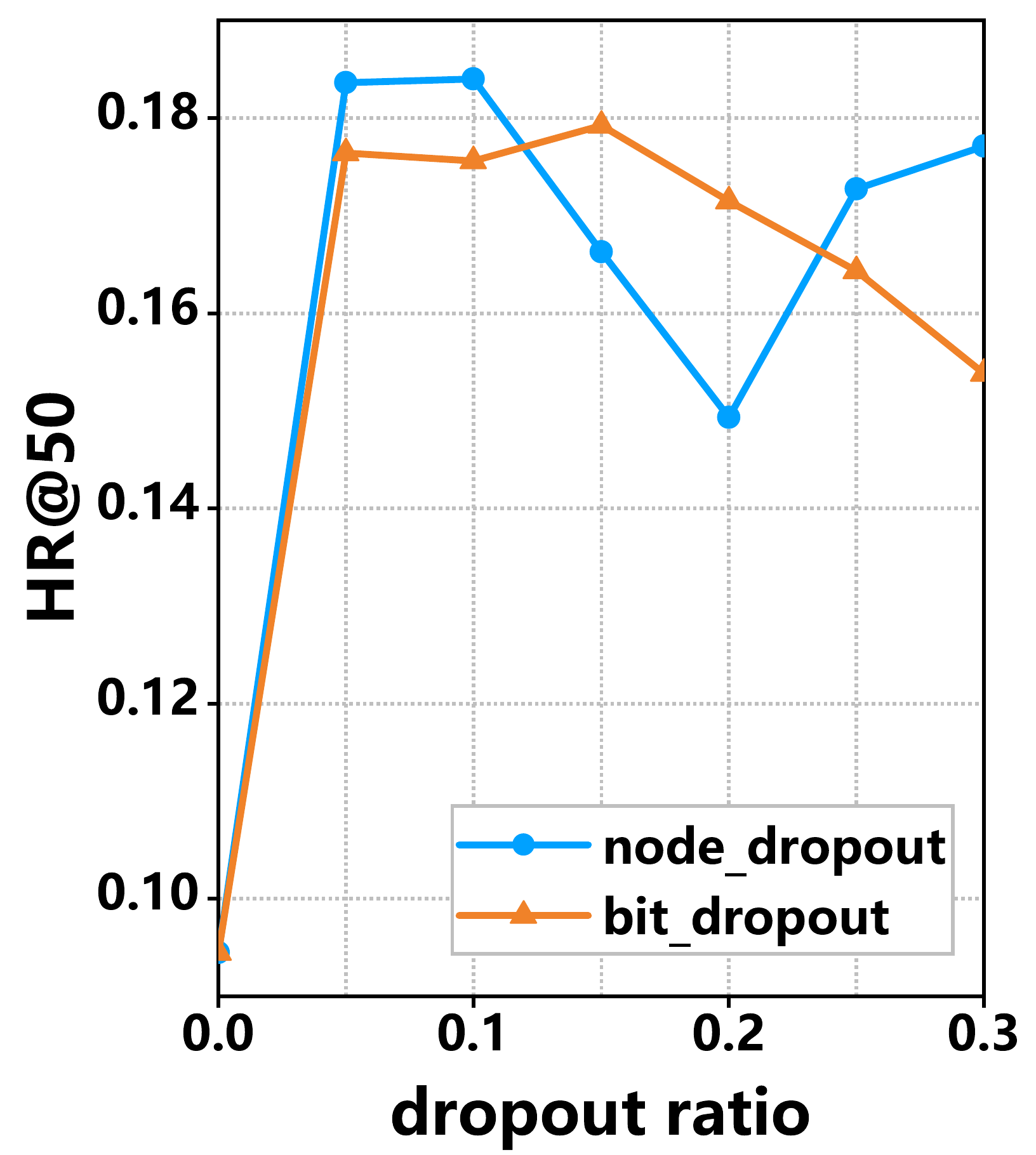}
\centerline{(a)~MovieLens}
\end{minipage}%
\begin{minipage}[t]{0.33\linewidth}
\centering
\includegraphics[width=\linewidth]{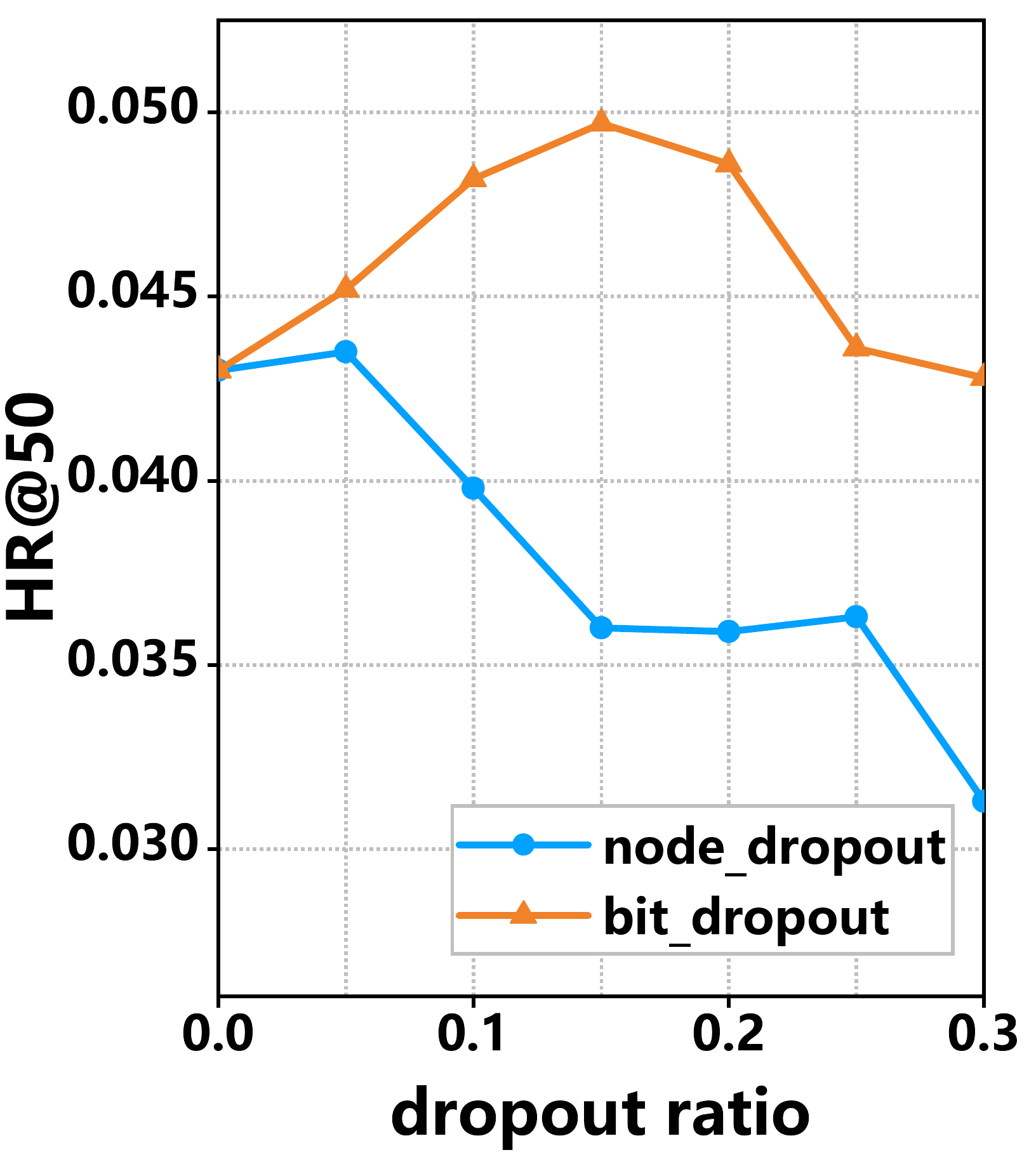}
\centerline{(b)~Yelp}
\end{minipage}%
\begin{minipage}[t]{0.33\linewidth}
\centering
\includegraphics[width=\linewidth]{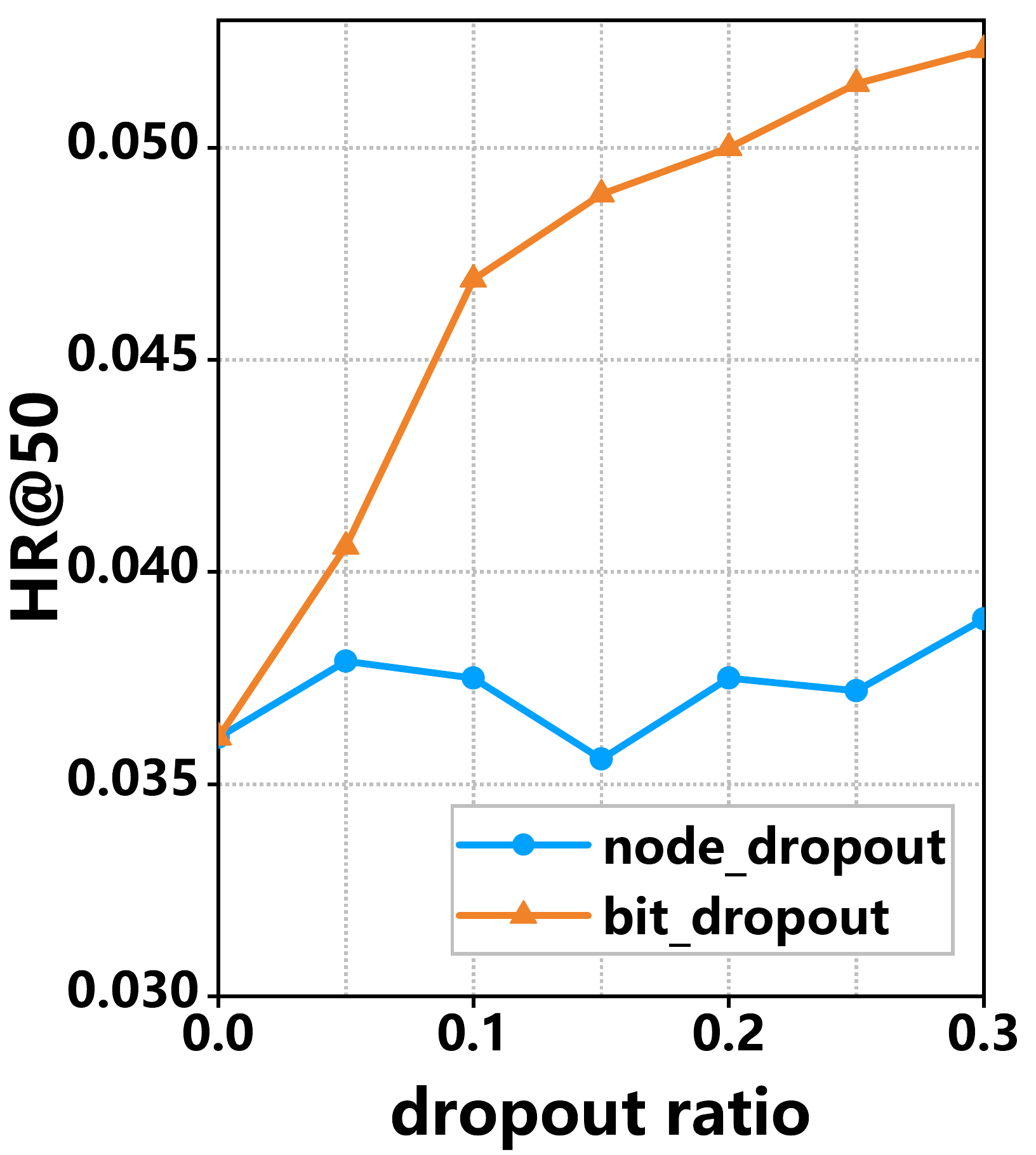}
\centerline{(c)~Amazon}
\end{minipage}%
\centering
\caption{Effect of node dropout and bit dropout ratios.}
\label{fig:drop}
\end{figure}

\begin{figure}[tbp]
\centering
\begin{minipage}[t]{0.495\linewidth}
\centering
\includegraphics[width=\linewidth]{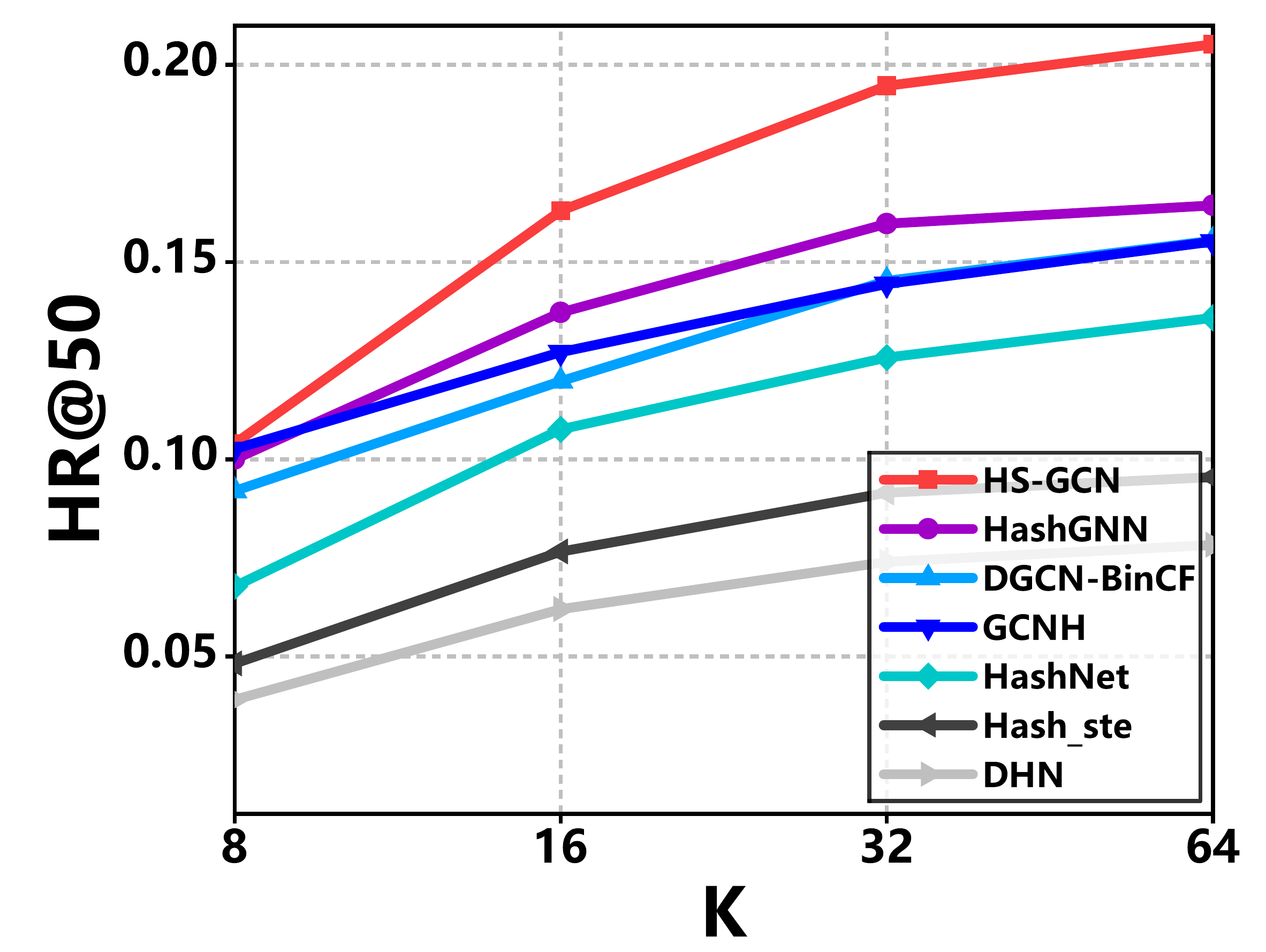}
\centerline{(a)~MovieLens}
\end{minipage}
\begin{minipage}[t]{0.495\linewidth}
\centering
\includegraphics[width=\linewidth]{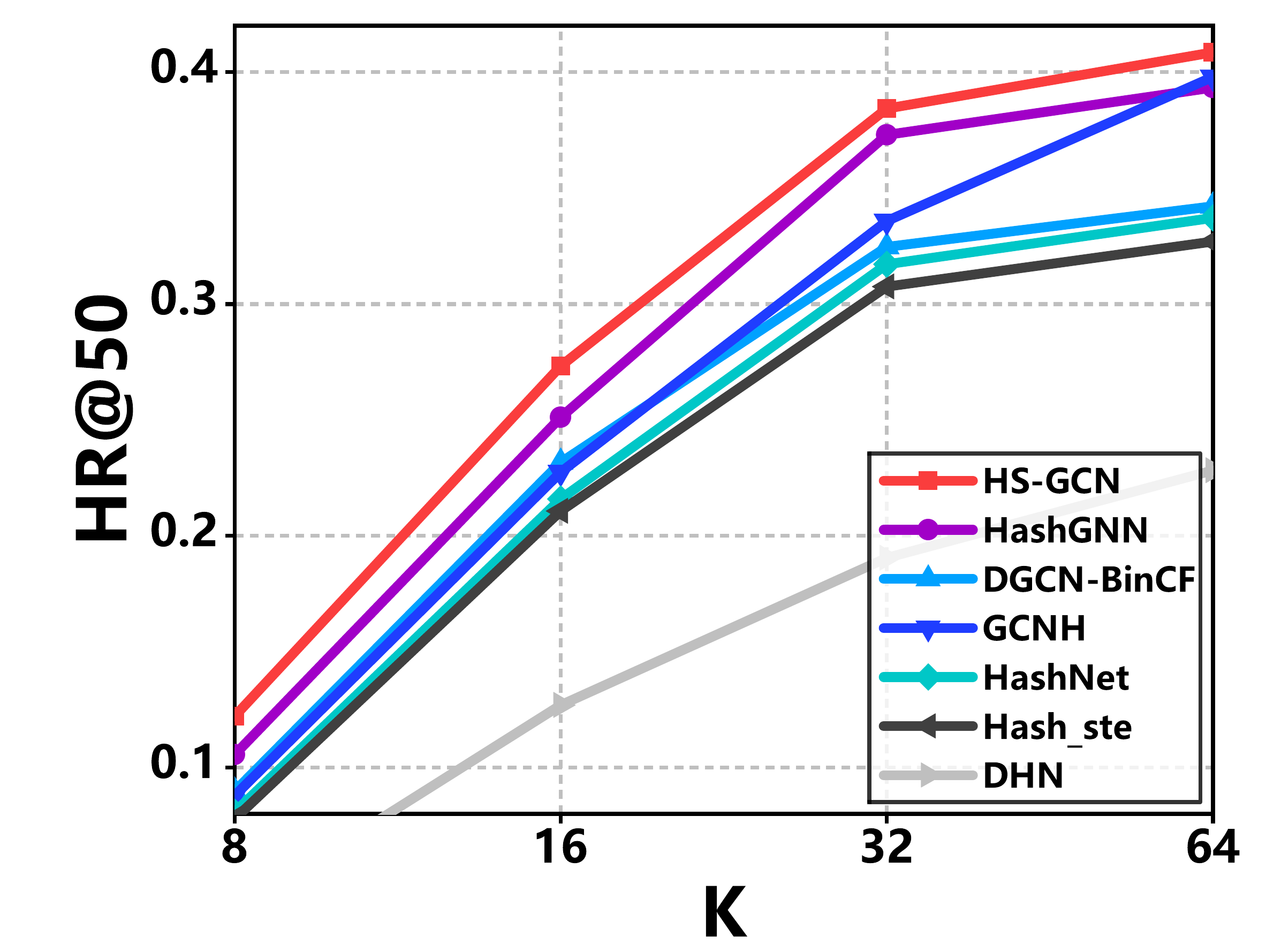}
\centerline{(b)~Gowalla}
\end{minipage}
\centering
\caption{Effect of the length of hash codes.}
\label{fig:length_effect}
\end{figure}

\subsubsection{Effect of Dropout}
Although GCNs have strong representation ability, they usually suffer from the overfitting problem. Dropout is an effective solution to prevent models from overfitting~\cite{srivastava2014dropout}. Following the prior GCN-based researches and considering our model architecture, we attempted to employ two dropout techniques: node dropout and bit dropout to improve the performance of HS-GCN. Node dropout randomly discards particular neighbor nodes during the aggregation of hash codes. For the $l$-th propagation layer, we randomly dropped $p_1\times (N+M)$ nodes of the adjacency matrix, where $p_1$ is the dropout ratio. We also conducted bit dropout that randomly drops a few bits of the final user and item hash codes before the prediction, with a probability $p_2$. 

Note that dropout is only used during the training process, and should be disabled in the testing process. Figure $\ref{fig:drop}$ plots the effect of node dropout ratio $p_1$ and bit dropout ratio $p_2$ under HR@50 evaluation metric on different datasets. Regarding the two dropout strategies, bit dropout delivers much better performance especially on sparse datasets. Node dropout even has a negative effect on Yelp dataset. One reason might be that dropping neighbor nodes damages the hash code aggregation operation in the propagation process of HS-GCN, especially when the neighbor nodes are sparse. Bit dropout is more effective, which means that it can be an effective strategy to address the overfitting of learning to hash on GCNs.

\subsubsection{Effect of the Length of Hash Codes}
In order to explore how the length of hash code affects HS-GCN and other baselines, we changed the length of their hash codes within $\{8,16,32,64\}$, and repeated the experiments. Figure~\ref{fig:length_effect} shows the performance of these models with different length $K$ on MovieLens and Gowalla datasets. It can be observed that the hash code length has positive influence on hashing models, while the influence weakens as length $K$ increases. Similar trends are observed on other datasets.

%% file: 6_conclusion.tex
\section{Conclusion and Future Work}
In this work, we explicitly incorporate the high-order Hamming similarity into the hash code learning. Particularly, we devise a novel framework HS-GCN, which yields binary hash codes by propagating the similarity information on the user-item graph structure. HS-GCN is the first proposed graph convolutional network in the Hamming space, where the propagation layers are built upon Hamming space operations to directly capture the high-order Hamming similarity. Extensive experiments on three real-world datasets demonstrate the rationality and effectiveness of capturing the high-order similarity for learning to hash. 

In future, we plan to strengthen HS-GCN by incorporating multi-form knowledge~\cite{liu2019maml}, like attributes in tables~\cite{wang2019kgat}, and celebrities in matrices~\cite{song2018neural,sun2019supervised}. Moreover, we are interested in building recommender systems for micro videos~\cite{wei2019neural}. Considering the micro videos constantly emerge in large numbers, they thus require more efficient and accurate recommendation. Another emerging research direction is to explore the interpretable hashing-based recommendation~\cite{zhang2019ears}. 